\documentclass{cernyrep}
\usepackage{texnames}
\usepackage[T1]{fontenc}
\usepackage{stfloats}
\def\wake{W}							
\def\Np{N}								
\def\gammarel{\gamma}					
\def\BBimpedanceVertical{Z_y^{\textrm{B\!B}}}		
\def\ResonatorFrequency{\omega_r}		
\def\QualityFactor{Q}						
\def\shuntImpedance{R_s}				
\def\blength{\tau}				
\def\emitlong{\varepsilon_l}					
\def\omegarev{\omega_0}				
\usepackage{varwidth}
\usepackage{xcolor}

\begin{document}
\title{Beam Instabilities}

\author{G.~Rumolo}

\institute{CERN, Geneva, Switzerland}

\maketitle 

\begin{abstract}

When a beam propagates in an accelerator, it interacts with both the external
fields and the self-generated electromagnetic fields. If the latter are
strong enough, the interplay between them and a perturbation in the beam
distribution function can lead to an enhancement of the initial perturbation,
resulting in what we call a beam instability. This unstable motion can
be controlled with a feedback system, if available,
or it grows, causing beam degradation and loss.
Beam instabilities in particle accelerators have been
studied and analysed in detail since the late 1950s. The subject owes its relevance to the fact
that the onset of instabilities usually determines the performance of an
accelerator. Understanding and suppressing the underlying sources and
mechanisms is therefore the key to overcoming intensity
limitations, thereby pushing forward the performance reach of a machine.
\end{abstract}

\section{Introduction}
\begin{figure}[htb]
\begin{center}
\includegraphics[trim = 25mm 35mm 25mm 35mm, clip, width=0.8\textwidth]{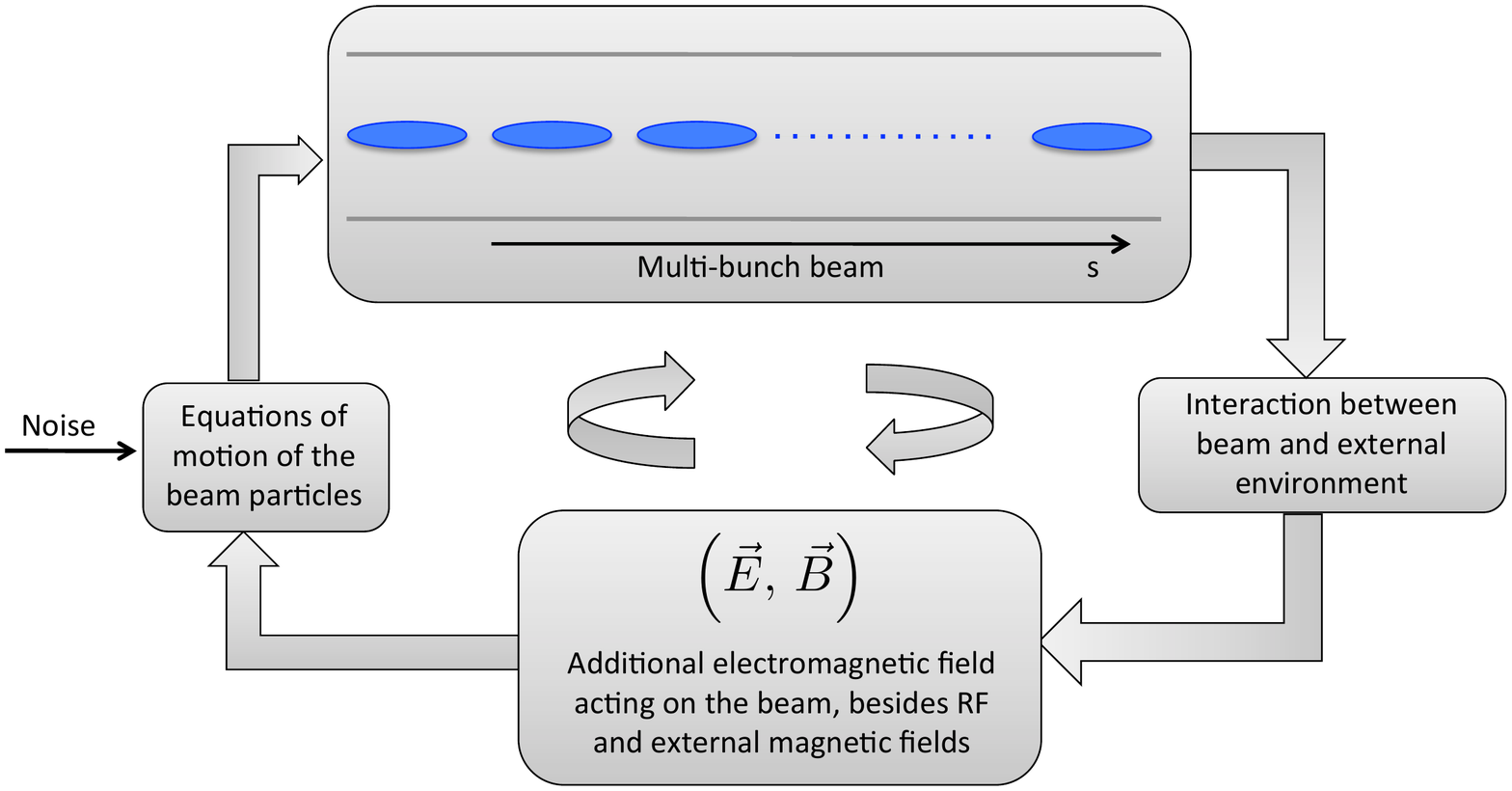}
\caption{Schematic of the closed loop through which a beam can become unstable under
the effect of self-generated electromagnetic fields.}
\label{instab-loop}
\end{center}
\end{figure}

The motion of charged particles forming a beam in an accelerator can be studied either individually or taking into account the electromagnetic interaction
between them.
In the former case, the beam is regarded as a collection of non-interacting
particles and the forces acting on them, i.e.\ the driving terms in each
particle's equations of motion, are fully prescribed by the
accelerator design. The study of the single-particle dynamics is then
complicated by all non-linear components of the applied electromagnetic fields.
In practice, this description is sufficient as long as additional electromagnetic
fields caused by the presence of the whole beam of particles are not strong enough
to perturb significantly the motion imparted by the external fields.
In many applications, however, beams carrying a high charge
(high intensity) and densely packed in a tiny phase space (high brightness)
are required, for which the electromagnetic fields created
by the interaction of the beam with the external environment need to be included when solving the particles' motion.
Under unfavourable conditions, these electromagnetic fields act back on the beam distribution
itself in a closed loop, such as to enhance a however small initial perturbation.
This situation eventually leads to an instability. The most general example of an instability loop is schematically
illustrated in Fig.~\ref{instab-loop}.

The block labelled `Interaction between beam and external environment' has been
willingly left vague, as any further specification depends on the type of problem
being modelled. In the most frequent case, which will also be the subject of this article, the interaction between beam and external
environment will be purely electromagnetic, so that it can be expressed in terms
of Maxwell's equations with the beam as source term and boundary conditions given
by the accelerator devices through which the beam propagates. Another case that is frequently the object
of study is when the beam generates an
electron or ion cloud that acts back on the beam itself and potentially destabilizes it. In this case, the interaction of
the beam with the environment needs to be described with all of the physical processes
leading to the cloud formation. The additional electric field from the cloud can then
be evaluated through Poisson's equation and used as a driving term in the equations
of motion of the beam particles.

\begin{figure}[htb]
\begin{center}
\includegraphics[trim = 0mm 0mm 0mm 8mm, clip, width=0.8\textwidth]{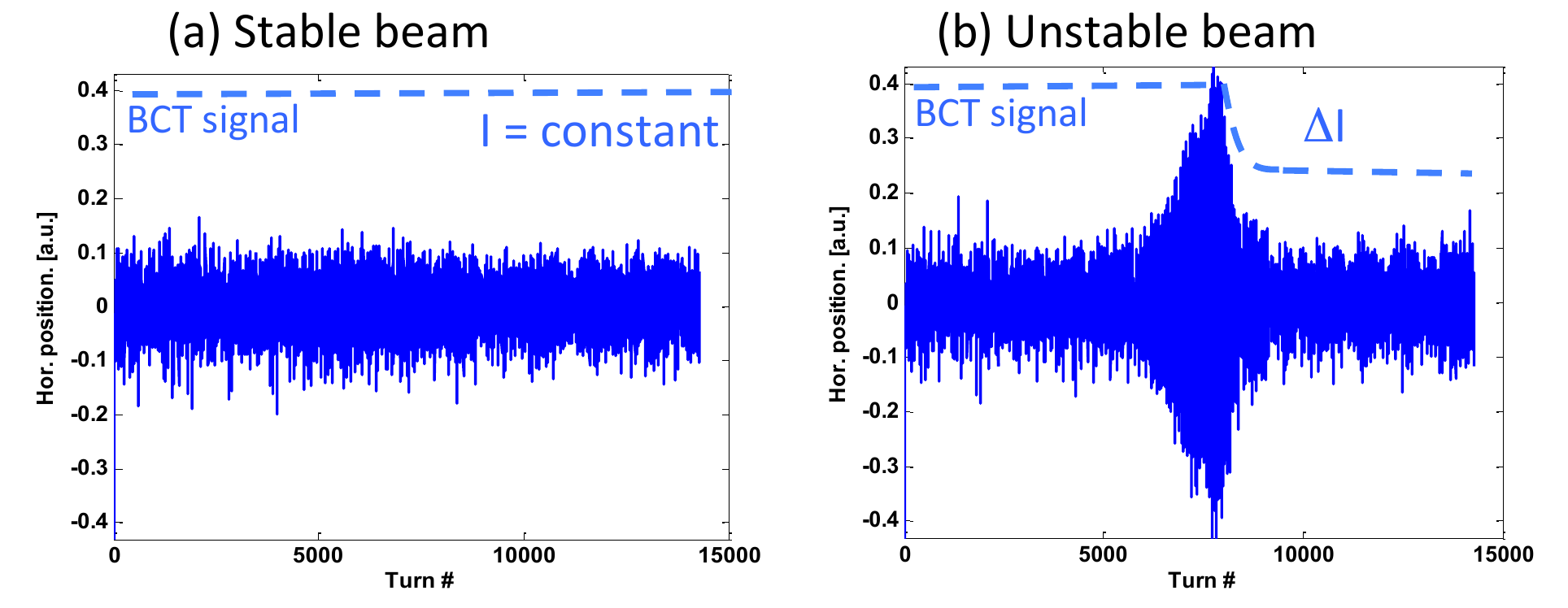}\\
{\small (a) \hspace{14pc} (b)}
\caption{Examples of stable (a) and unstable (b) signals from a BPM. The
signal from a beam current transformer (BCT) is also sketched, showing how the stable
beam does not suffer from any intensity loss, while a sharp intensity decrease is associated
with the rise of the instability.}
\label{instability}
\end{center}
\end{figure}

In practice, a beam becomes unstable when, as a result of the loop described above, at least a moment of its six-dimensional (6D) phase space
distribution, $\psi(x,y,z,x',y',\delta)$, exhibits an exponential growth (e.g.\ typically the
mean positions $\langle x\rangle$,  $\langle y\rangle$,
$\langle z\rangle$ or the standard deviations $\sigma_x$, $\sigma_y$, $\sigma_z$),
resulting in beam loss or emittance growth. Assuming an arbitrary observation point $s_0$ along the trajectory
of a beam inside an accelerator, described through the coordinate $s$, the full 6D phase space can usually be decomposed into transverse and longitudinal phase spaces. The 4D transverse space is described by the two pairs of conjugate variables $(x, x',y, y')$, i.e.\ the offsets from the nominal orbit in the horizontal and vertical directions (horizontal is the direction in which the beam is bent), and the relative divergences from the nominal orbit, $x'={\rm d}x/{\rm d}s$ and $y'={\rm d}y/{\rm d}s$. The longitudinal plane is described by the conjugate pair $(z,\delta)$, i.e.\ a space coordinate proportional to the delay in the arrival time at the selected location with respect to the synchronous particle, $z=-c\tau$ (the minus sign is chosen such that particles arriving before the synchronous particle have a positive $z$), and the relative longitudinal momentum deviation from the nominal momentum, $\delta=\delta p/p_0$. As an example of instability detection, the onset of a transverse instability can be easily revealed by the signal captured from a beam position monitor (BPM). A phase of exponential growth can be observed, usually followed by saturation and decay either due to non-linearities or because of beam loss. Figure \ref{instability} shows an example of horizontal BPM signals from two different bunches during the store of a train of 72 bunches with 25~ns spacing in the CERN-Proton Synchrotron (PS). The signal in Fig.~\ref{instability}(a) is basically BPM noise and represents a stable bunch, while that in Fig.~\ref{instability} is from an unstable one. This also highlights how the unstable beam oscillation is eventually associated with a certain amount of beam loss and is damped after the loss occurs.

The interest in studying coherent beam instabilities arises from the fact that the onset of a beam instability usually determines the maximum beam intensity that a machine can store/accelerate (i.e.~its performance limitation). Understanding the type of instability limiting the performance, and its underlying mechanism, is essential because it allows the source and possible measures to mitigate/suppress the effect to be identified, or providing the specifications of an active feedback system to prevent the instability. Beam instabilities occur in both linear and circular machines and can equally affect the longitudinal plane or the transverse plane. Coherent instabilities can affect the beam on different scales. For example, a typical multibunch instability exhibits an excitation pattern extending over different bunches in a train and depends on a long-range coupling agent. Nevertheless, in some cases the unstable motion of subsequent bunches does not appear as coupled, because the instability can be just the consequence of a certain mechanism that builds up along the bunch train, but visibly affects only the last bunches of a train (e.g.\ an electron cloud). In a pure single-bunch instability, usually the coupling happens between head and tail of the same bunch. In this case, the mechanism that drives the instability only needs to act on the short range.

In the following sections, we will first set the mathematical framework to address the problem of beam instabilities driven by self-generated electromagnetic fields (wake functions and impedances) and we will then apply these concepts to reduced models (one- or two-particle) to explain the physics of some of the most frequent instability mechanisms in particle accelerators. The reference that will be followed throughout this article is \cite{Chao}.

\section{The longitudinal plane}
\begin{figure}[htb]
\begin{center}
\includegraphics[trim = 10mm 76mm 80mm 45mm, clip, width=0.8\textwidth]{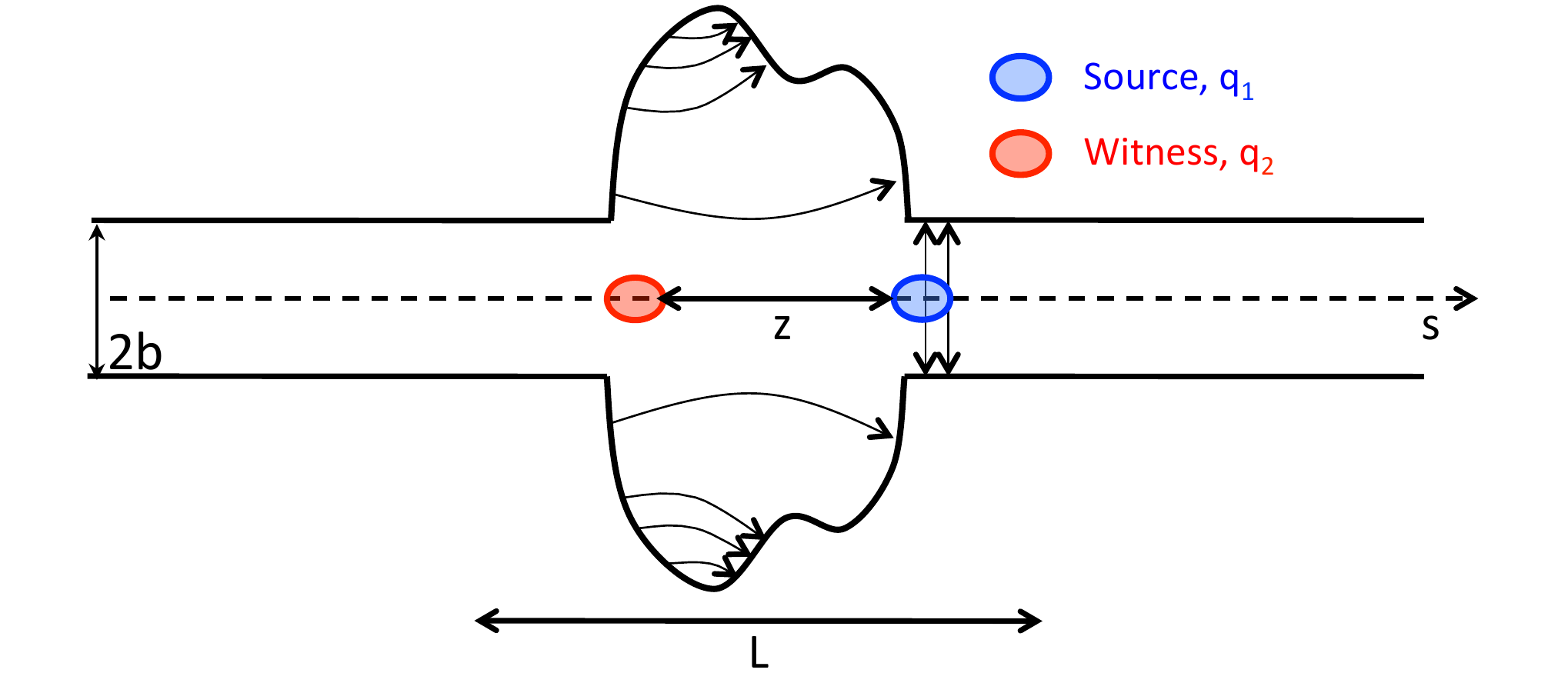}
\caption{Wake field from a source particle potentially affecting a witness travelling at distance $z$ behind the source}
\label{sketch-wake}
\end{center}
\end{figure}

Let us consider two ultra-relativistic charged particles ($q_1$ and $q_2$, travelling at $v\approx c$, or equivalently $\gamma\gg 1$) going through an accelerator structure, separated by a distance $|z|$ ($z=-c\tau$, with $\tau$ expressing the delay between the arrival times of the two particles at an arbitrary location). The leading particle will be our source and the trailing particle will be the witness. Since both particles are travelling basically at the speed of light, causality imposes that the leading particle cannot be affected by the trailing particle. As long as source and witness move in a perfectly conducting chamber, the witness does not feel any force from the source. However, when the source encounters a discontinuity, the electromagnetic field produced to satisfy the boundary conditions (wake field) can effectively reach the witness particle and affect its motion. In this process, the source loses energy, while the witness feels a net force all along an effective length, $L$, of the discontinuity/structure/device that caused the wake. Figure \ref{sketch-wake} shows a simple sketch of how the situation could look like after a source has gone through a cavity-like object and modes are trapped after its passage.

In fact, geometric discontinuities are not the only possible origin of wake fields. For example, in a chamber with finite conductivity the induced current from a source particle is delayed and can significantly act back on witness particles within a certain distance range. Generally, electromagnetic boundary conditions other than a perfect electrical conductor (PEC) can generate wake fields.

The {\em longitudinal wake function} associated with a certain accelerator object (able to create a wake field) is defined as the integrated longitudinal force ($q_2E_s(s,z)$) acting on the witness particle along the effective length $L$ of the object (i.e.\ its energy change, $\Delta E_2$), normalized by the source and witness charges:
\begin{equation}
W_{||}(z)=-\frac{\int_0^L E_s(s,z) \, {\rm d}s}{q_1} = -\frac{\Delta E_2}{q_1q_2}.
\label{longwake}
\end{equation}

The minus sign is also introduced in the definition, so that $W(0)=-\Delta E_1/q_1^2$ is defined positive (the source particle can only lose energy, $\Delta E_1<0$). The beam loading theorem also proves that the wake function is discontinuous in $z=0$, with $W_{||}(0^-)=2\cdot W_{||}(0)$. Intuitively, this theorem states that a particle travelling at the speed of light can only see half of its own wake. Besides, causality imposes that $W_{||}(0^+)=0$, and actually $W_{||}(z)=0$ for $z>0$. In a global energy balance, the energy lost by the source, $\Delta E_1$, splits into
\begin{itemize}
\item Electromagnetic energy of the modes that may remain trapped in the object. This is then partly dissipated on the lossy walls or into purposely designed inserts or higher order mode (HOM) absorbers. Partly, it can be potentially transferred to the trailing particles (or the same particle over successive turns), possibly feeding into an instability.
\item Electromagnetic energy of modes that propagate down the beam chamber (above cut-off), which will be eventually lost on surrounding lossy materials.
\end{itemize}
The energy loss of a beam is very important, because the fraction lost on the beam environment causes equipment heating (with consequent outgassing and possible damage), while the part associated with long-lived wake fields can feed into both longitudinal and transverse instabilities. The calculation of the energy loss will be the subject of the next subsection.

The wake function of an accelerator object is basically its Green function in the time domain (i.e. the electromagnetic response of the object to a pulse excitation). Therefore, it is very useful for macroparticle models and simulations, because it can be used to describe the driving terms in the single-particle equations of motion, as we will see in one of the next subsections. However, we can also describe this response as a transfer function in the frequency domain.
This gives the definition of {\em longitudinal beam coupling impedance} of the object under study:
\begin{equation}
Z_{||}(\omega)=\int_{-\infty}^{\infty} W_{||}(z) \exp\left(-\frac{{\rm i}\omega z}{c}\right)\frac{{\rm d}z}{c}.
\label{longimped}
\end{equation}

\begin{figure}[htb]
\begin{center}
\includegraphics[trim = 20mm 20mm 20mm 20mm, clip, width=0.8\textwidth]{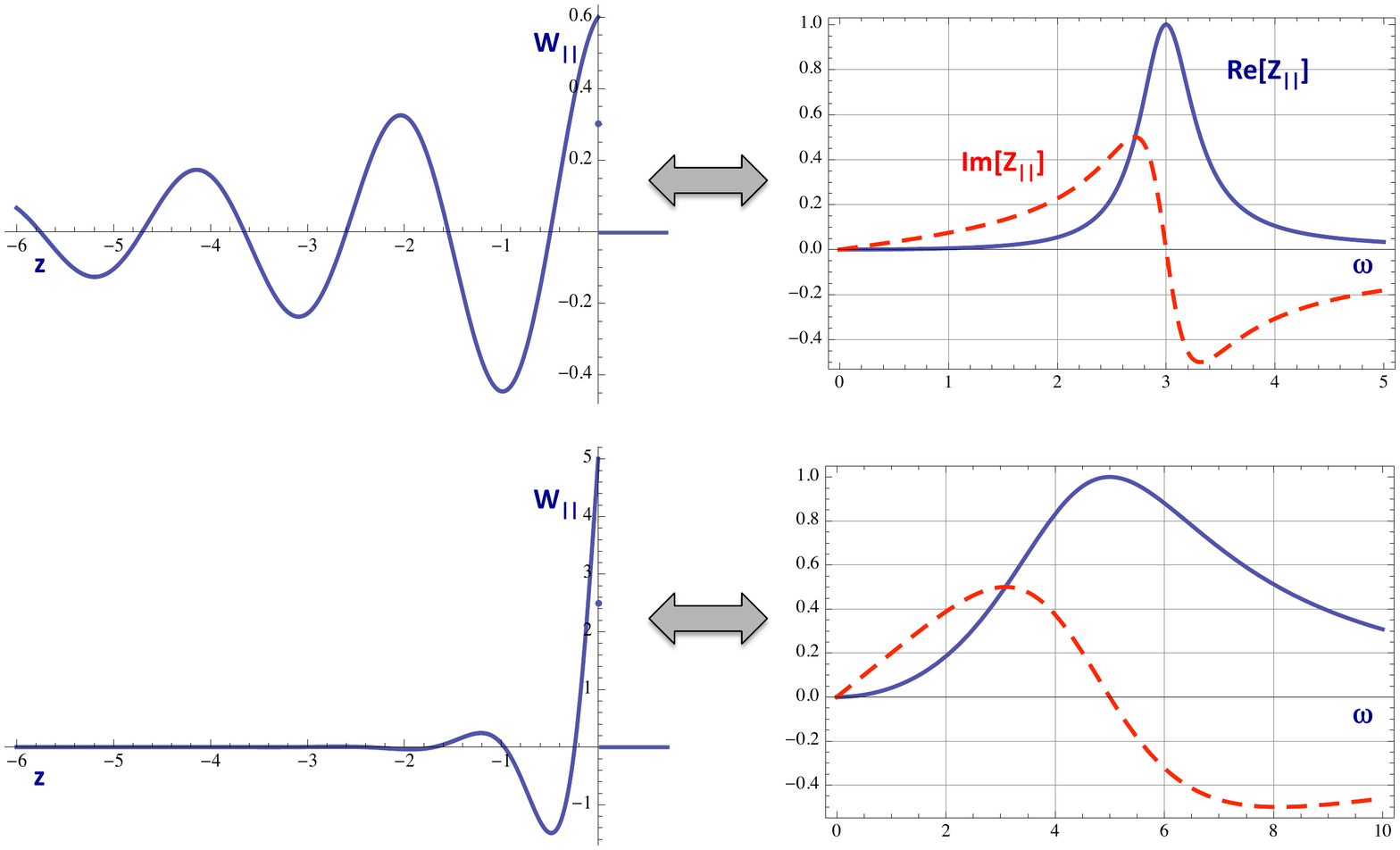}
\caption{Wake functions (left) and beam coupling impedances (right) for narrowband (top) and broadband (bottom) resonator objects.}
\label{resonators}
\end{center}
\end{figure}

\begin{figure}[htb]
\begin{center}
\includegraphics[trim = 30mm 60mm 25mm 60mm, clip, width=0.8\textwidth]{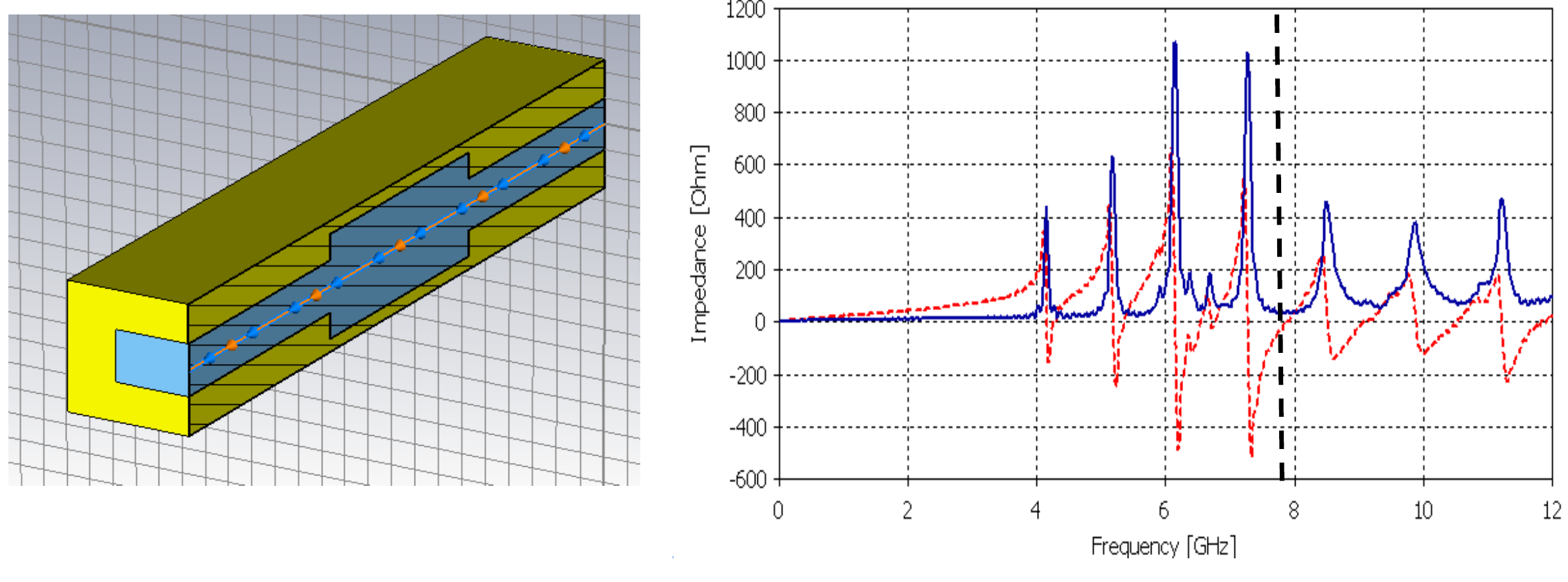}\\
{\small (a) \hspace{14pc} (b)}
\caption{Pill-box cavity: a 3D longitudinal cut of the simulated cavity (a) and the obtained
longitudinal beam coupling impedance (b). The cut-off frequency of the beam chamber is shown with a vertical dashed line. Courtesy of C.~Zannini.}
\label{pillbox}
\end{center}
\end{figure}

Typical longitudinal wake/impedance pairs are described as resonators and are displayed in Fig.~\ref{resonators}. The wake function is a damped oscillation with a discontinuity in $z=0$, while the beam coupling impedance spectrum exhibits a peak at the specific oscillation frequency. The width of the peak relates to the lifetime of the oscillation in the time domain before becoming fully damped, distinguishing between  a narrowband and a broadband resonator, as shown in top and bottom of Fig.~\ref{resonators}, respectively. In more complex cases, several modes can be excited in the object and the beam coupling impedance is a combination of several peaks like those shown in the single-resonance examples depicted in Fig.~\ref{resonators}. For example, a pill-box cavity with walls having finite conductivity and attached to a vacuum chamber left and right (Fig.~\ref{pillbox}(a)) can resonate on all its characteristic modes determined by its geometry. The width of the excited peaks will be narrower for the modes below the cut-off frequency of the chamber (as the decay is purely determined by the resistive losses), while they will be broader for the peaks above cut-off, as additional losses come from the propagation of these modes into the chamber. This is visible in Fig.~\ref{pillbox} (simulations done with CST\textregistered \hspace{1mm} Particle Studio Suite).

In beam physics, broadband impedances, such that the associated wake functions decay over the length of one particle bunch, are only responsible for intrabunch (head--tail) coupling, potentially leading to single-bunch instabilities. Conversely, narrowband impedances, associated with long-lived wake functions decaying over the length of a bunch train or several turns, cause bunch-to-bunch or multiturn coupling, leading to multibunch or multiturn instabilities.

\subsection{Energy loss}
\begin{figure}[htb]
\begin{center}
\includegraphics[trim = 70mm 80mm 90mm 30mm, clip, width=0.6\textwidth]{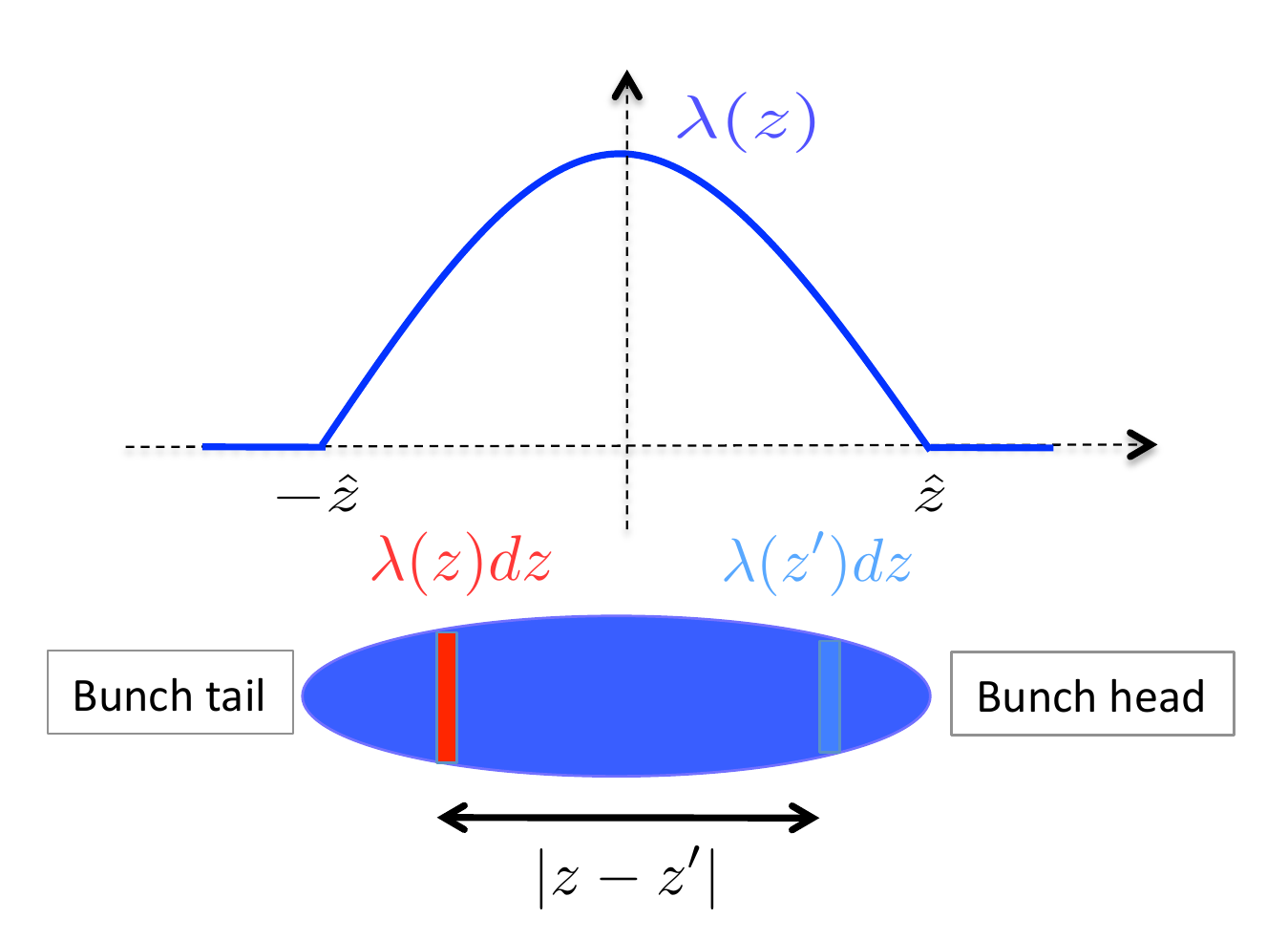}
\caption{Sketch of the bunch and line density. Source and witness slices are also highlighted}
\label{lambda}
\end{center}
\end{figure}

By using the concepts so far introduced, we can easily derive an analytical expression for the energy lost by a bunch with line density $\lambda(z)$ (see Fig.~\ref{lambda}) when it goes through a structure characterized by a wake function $W_{||}(z)$ or beam coupling impedance $Z_{||}(\omega)$. The energy change $\Delta E(z)$ of the witness slice $e\lambda(z) \, {\rm d}z$ can be expressed as the integral of the contributions from the wake functions generated by all the preceding source slices, $e\lambda(z') \, {\rm d}z'$. Integrating $\Delta E(z)$ over the whole bunch provides the total energy loss of the bunch:
\begin{equation}
\Delta E = \int_{-\hat{z}}^{\hat{z}}\Delta E(z) \, {\rm d}z =  - \int_{-\hat{z}}^{\hat{z}} e\lambda(z) \int_z^{\hat{z}}e\lambda(z')W_{||}(z-z') \, {\rm d}z' \, {\rm d}z.
\label{bunchenloss1}
\end{equation}

By using the Parseval identity and the convolution theorem, the energy loss can be easily written in terms of bunch spectrum $\Lambda(\omega)$ and beam coupling impedance:
\begin{equation}
\Delta E = -\frac{e^2}{2\pi}\int_{-\infty}^{\infty}\Lambda^*(\omega)\left[\Lambda(\omega)Z_{||}(\omega)\right] \, {\rm d}\omega=-\frac{e^2}{2\pi}\int_{-\infty}^{\infty}|\Lambda(\omega)|^2\mathrm{Re}\left[Z_{||}(\omega)\right] \, {\rm d}\omega.
\label{bunchenloss2}
\end{equation}

In the last expression, we also took into account that, since $W_{||}(z)$ is a real function, $\mathrm{Re}[Z_{||}(\omega)]$ and $\mathrm{Im}[Z_{||}(\omega)]$ are even and odd functions of $\omega$, respectively. Since Eq.~(\ref{bunchenloss2}) represents the total energy lost by the bunch over a single pass through the object with beam coupling impedance $Z_{||}(\omega)$, it can also be interpreted as the bunch energy loss per turn in a circular accelerator (again due to a single object with beam coupling impedance $Z_{||}(\omega)$, or the total energy loss per turn if $Z_{||}(\omega)$ represents instead the total longitudinal beam coupling impedance modelling the whole ring). However, this is rigorously true only as long as the wake function is short enough lived to be fully damped after one turn, so that subsequent passages of the bunch are not coupled through the wake.

In fact, defining $C$ as the circumference of the ring, Eqs.~(\ref{bunchenloss1}) and (\ref{bunchenloss2}) can be generalized to the case of a bunch going through a structure that keeps memory of previous passages, assuming that its longitudinal distribution does not change in time:
\begin{equation}
\Delta E = \int_{-\hat{z}}^{\hat{z}}\Delta E(z) \, {\rm d}z =  - \int_{-\hat{z}}^{\hat{z}} e\lambda(z) \int_z^{\hat{z}}e\lambda(z')\sum_{k=-\infty}^{\infty}W_{||}(kC+z-z') \, {\rm d}z' \, {\rm d}z.
\label{bunchenloss1multi}
\end{equation}

Applying the identity
\begin{equation}
\sum_{k=-\infty}^{\infty}W_{||}(kC+z-z')=\frac{\omega_0}{2\pi}\sum_{p=-\infty}^{\infty}Z_{||}(p\omega_0)\exp\left[-\frac{{\rm i}p\omega_0(z-z')}{c}\right],
\end{equation}
\noindent in which $\omega_0=2\pi c/C$ is the revolution frequency, we can easily recast Eq.~(\ref{bunchenloss1multi}) in the following form:
\begin{equation}
\Delta E = -\frac{e^2\omega_0}{2\pi}\sum_{p=-\infty}^{\infty}|\Lambda(p\omega_0)|^2\mathrm{Re}\left[Z_{||}(p\omega_0)\right].
\label{bunchenloss2multi}
\end{equation}

Equation (\ref{bunchenloss2multi}) is very powerful, because it can be applied to the full beam circulating in an accelerator ring and can be used for calculating the total beam energy loss per turn. In this case, we would simply need to replace $\Lambda(\omega)$, the Fourier transform of the single-bunch distribution, with the Fourier transform of the full beam signal, $\Lambda_B(\omega)$. For example, we could assume the beam to be a train of $M$ bunches covering only a fraction of the full circumference ($M<h$, $h$ being the harmonic number of the accelerator) with spacing between bunches $\tau_b=2\pi/(h\omega_0)$:
\begin{equation}
\lambda_B(z)=\sum_{n=0}^{M-1}\lambda(z-nc\tau_b)  \;\; \stackrel{\displaystyle\mathcal{F}}{\iff} \;\; \Lambda_B(\omega)=\Lambda(\omega)\sum_{n=0}^{M-1}\exp\left(-{\rm i}\omega\tau_b\right).
\end{equation}

Summing up the terms in the expression of the beam spectrum, we obtain
\begin{equation}
\Lambda_B(\omega)=\Lambda(\omega)\exp\left[\-\frac{{\rm i}\omega\tau_b(M-1)}{2}\right]\cdot\frac{\displaystyle\sin\left(\frac{M\omega\tau_b}{2}\right)}{\displaystyle\sin\left(\frac{\omega\tau_b}{2}\right)},
\end{equation}
\noindent which can be finally inserted into Eq.~(\ref{bunchenloss2multi}), yielding
\begin{equation}
\Delta E = \frac{e^2\omega_0}{2\pi}\sum_{p=-\infty}^{\infty}|\Lambda(p\omega_0)|^2\mathrm{Re}\left[Z_{||}(p\omega_0)\right]
\cdot \left[\frac{1-\displaystyle\cos\left(\frac{2\pi Mp}{h}\right)}{1-\displaystyle\cos\left(\frac{2\pi p}{h}\right)}\right].
\end{equation}

The terms in the summation above are maximum for $p=k\cdot h$, as the ratio in brackets becomes equal to $M^2$. This means that narrowband impedances peaked around multiples of the harmonic number of the accelerator are the most efficient to drain energy from the beam, and consequently the associated objects suffer from beam-induced heating. However, this type of impedances, usually associated with the RF systems and their HOMs, need to be avoided in accelerator design by either detuning them or including HOM absorbers. In fact, they not only cause equipment heating, but potentially lead to important instabilities (e.g.\ the Robinson instability, see the next subsection, or transverse coupled bunch instabilities).

The total energy loss per turn associated with the global accelerator impedance needs to be compensated for by the RF system, so that the average stable phase shifts by an amount $\langle\Delta \Phi_s\rangle$ given by
\begin{equation}
\sin\langle\Delta\Phi_s\rangle = \frac{\Delta E}{MN_beV_m},
\label{phaseshift}
\end{equation}

\noindent where $N_b$ is the number of particles per bunch and $V_m$ is the applied RF voltage.

\subsection{The Robinson instability}
To study instabilities, the effect of wake fields (or impedances) must be formally introduced in the equation of motion of the beam particles. Resorting to the concepts introduced at the beginning of this section, we can write the equation of motion of any single particle in the witness slice $\lambda(z) \, {\rm d}z$ under the effect of the force from the RF system and that associated with the wake, which can extend to several turns:
\begin{equation}
\frac{{\rm d}^2z}{{\rm d}t^2} + \frac{\eta e V_{\mathrm{RF}}(z)}{m_0\gamma C} = \frac{\eta e^2}{m_0\gamma C}\int_{z}^{\infty}\sum_{k=0}^{\infty}\lambda(z'+kC, t)W_{||}(z-z'-kC) \, {\rm d}z'.
\label{motion}
\end{equation}

Equation (\ref{motion}) is very general and can be used in macroparticle tracking programs, which solve it for each macroparticle, determining self consistently the full beam evolution $\lambda(z,t)$. It is to be noted that both the integral and the summation in the above equation can be formally extended from $-\infty$, as the wake function vanishes for positive values of $z$ due to causality.

In the following, to illustrate the most basic mechanism of longitudinal instability, i.e. the {\em Robinson instability}, we will make use of these simplifications:
\begin{itemize}
\item The bunch is assumed to be point-like (carrying the full bunch charge $N_be$) and feels an external linear focusing force (i.e. in absence of the wake forces, it would execute linear synchrotron oscillations with synchrotron frequency $\omega_s$).
\item The bunch additionally feels the effect of the multiturn wake from an impedance source distributed over the ring circumference $C$ (the analysis would not change if the impedance source had been lumped at one ring location, and in reality both the external voltage and the impedance source should be localized, making Eq.~(\ref{motion}) {\em de facto} time discrete).
\end{itemize}
In this case, the equation of motion (\ref{motion}) reduces to
\begin{equation}
\frac{{\rm d}^2z}{{\rm d}t^2} +\omega_s^2z= \frac{N_b\eta e^2}{m_0\gamma C}\sum_{k=0}^{\infty}W_{||}\left[z(t)-z(t-kT_0)-kC\right].
\label{Rob-motion}
\end{equation}

First of all, we assume that the wake function can be linearized on the scale of the synchrotron oscillation (i.e. the wake function does not exhibit abrupt changes over a half-bucket length):
\begin{equation}
W_{||}\left[z(t)-z(t-kT_0)-kC\right]\approx W_{||}(kC)+W'_{||}(kC)\cdot\left[z(t)-z(t-kT_0)-kC\right].
\end{equation}

We can use the above expansion in Eq.~(\ref{Rob-motion}). The term  $\sum_k W_{||}(kC)$  only contributes to a constant term in the solution of the equation of motion, shifting the centre of the synchrotron oscillation from the bucket centre to a certain $z_0$. This term represents the stable phase shift that compensates for the energy loss introduced by the wake and will be neglected in the following. The dynamic term proportional to $z(t)-z(t-kT_0) \approx kT_0 \, {\rm d}z/{\rm d}t$, instead, is a friction-like term in the equation of the harmonic oscillator and, under certain conditions, can lead to an instability. Going to the frequency domain then yields
\begin{equation}
\omega^2 - \omega_s^2 = -\frac{{\rm i}N_b\eta e^2}{m_0\gamma C^2}\sum_{p=-\infty}^{\infty}\left[p\omega_0Z_{||}(p\omega_0)-(p\omega_0+\omega)Z_{||}(p\omega_0+\omega)\right].
\end{equation}

At this point, assuming that the wake only introduces a small deviation from the nominal synchrotron frequency, we can easily write the complex frequency shift, which results in a real part (synchrotron frequency shift) and an imaginary part (growth/damping rate):
\begin{equation}
\left\{\begin{array}{l}
\Delta\omega_s=\mathrm{Re}(\omega-\omega_s)=\displaystyle\left(\frac{e^2}{m_0c^2}\right)\frac{N_b\eta}{2\gamma T_0^2\omega_s}\sum_{p=-\infty}^{\infty}\left[p\omega_0\mathrm{Im}Z_{||}(p\omega_0)-(p\omega_0+\omega_s)\mathrm{Im}Z_{||}(p\omega_0+\omega_s)\right],\\[0.7cm]
\tau^{-1}=\mathrm{Im}(\omega-\omega_s)=\displaystyle\left(\frac{e^2}{m_0c^2}\right)\frac{N_b\eta}{2\gamma T_0^2\omega_s}\sum_{p=-\infty}^{\infty}(p\omega_0+\omega_s)\mathrm{Re}Z_{||}(p\omega_0+\omega_s).
\end{array}
\right.
\label{shifts}
\end{equation}
\vspace*{0.cm}

The possibility of having an instability is related to a positive value of $\tau$ in the second of the equations in~(\ref{shifts}). This is determined by the sign of $\eta$ and that of the weighted summation on $\mathrm{Re}Z_{||}$, which are the only two terms that can admit both signs.

A relevant situation that can be studied in further detail is when the impedance has a spectrum peaked at a frequency $\omega_r$ close to the RF frequency $h\omega_0$, or to a multiple of it (i.e. associated with the cavity fundamental mode or with a HOM). In this case, out of the infinite summation only two terms will dominate the right-hand side of the equation for the growth/damping rate:
\begin{equation}
\tau^{-1}=\mathrm{Im}(\omega-\omega_s)\approx\displaystyle\left(\frac{e^2}{m_0c^2}\right)\frac{N_b\eta h\omega_0}{2\gamma T_0^2\omega_s}\left[\mathrm{Re}Z_{||}(h\omega_0+\omega_s)-\mathrm{Re}Z_{||}(h\omega_0-\omega_s)\right].
\end{equation}

\begin{figure}[htb]
\begin{center}
\includegraphics[trim =25mm 50mm 20mm 40mm, clip, width=0.9\textwidth]{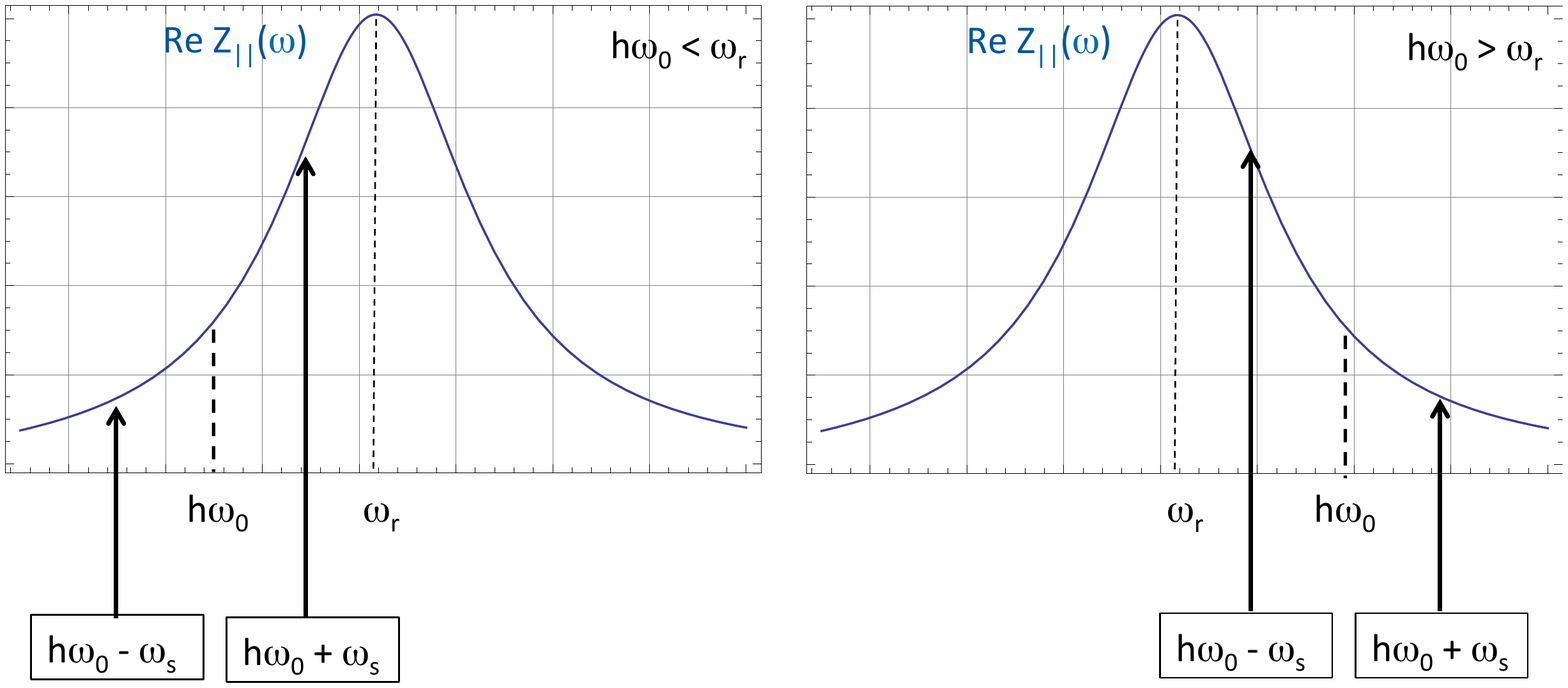}\\
{\small (a) \hspace{14pc} (b)}
\caption{Sketch of the two possible situations for the Robinson instability}
\label{rob-sketch}
\end{center}
\end{figure}

Stability requires that $\eta$ and the variation of $\mathrm{Re}Z_{||}(\omega)$ around $n\omega_0 $ have different signs. Figure \ref{rob-sketch} shows that, assuming $\omega_s$ to be small with respect to the width of the resonance peak, this can be achieved differently according to whether $\omega_r$ is below or above $n\omega_0$. In particular, when $h\omega_0<\omega_r$ (Fig.~\ref{rob-sketch}(a)), the term $\left[\mathrm{Re}Z_{||}(h\omega_0+\omega_s)-\mathrm{Re}Z_{||}(h\omega_0-\omega_s)\right]$ is positive and therefore $\eta$ needs to be negative for stability (i.e. the machine should be operating below transition). Otherwise, for $h\omega_0>\omega_r$ (Fig.~\ref{rob-sketch}(b)), stability is guaranteed only above transition.

Other types of impedances can also cause instabilities through the Robinson mechanism, following the general equations~(\ref{shifts}).
However, a smooth broadband impedance with no narrow structures on the $\omega_0$ scale cannot give rise to an instability, because
\begin{equation}
\sum_{p=-\infty}^{\infty}(p\omega_0+\omega_s)\mathrm{Re}Z_{||}(p\omega_0+\omega_s) \rightarrow \frac{1}{\omega_0}\int_{-\infty}^{\infty}\omega\mathrm{Re}Z_{||}(\omega) \, {\rm d}\omega \rightarrow 0.
\end{equation}

Physically, this could be expected, because the absence of structure on $\omega_0$ scale in the spectrum implies that the wake has fully decayed over one turn and, therefore, the driving term in the equation of motion (\ref{Rob-motion}) also vanishes.

To summarize, the Robinson instability affects a single bunch under the action of a multiturn wake field. It is characterized by a term of coherent synchrotron tune shift (the first of the equations~(\ref{shifts})) and an unstable rigid bunch dipole oscillation (growth rate given by the second of the equations~(\ref{shifts}) under the conditions explained above). It does not involve higher order moments of the bunch longitudinal phase space distribution. Other important collective effects can affect a bunch in a beam, for instance:
\begin{itemize}
\item Potential well distortion, resulting in synchronous phase shift, bunch lengthening or shortening, synchrotron tune shift/spread.
\item Coupled bunch instabilities.
\item Higher order mode and mode-coupling single-bunch instabilities (e.g. microwave instability).
\item Coasting beam instabilities (e.g. negative-mass instability).
\end{itemize}

To be able to study these effects, more refined modes of the beam are needed (e.g.\ the kinetic model described by the Vlasov equation or macroparticle simulations), but this is beyond the scope of this introductory article.

\section{The transverse plane}
\begin{figure}[htb]
\begin{center}
\includegraphics[trim = 10mm 10mm 80mm 10mm, clip, width=1.0\textwidth]{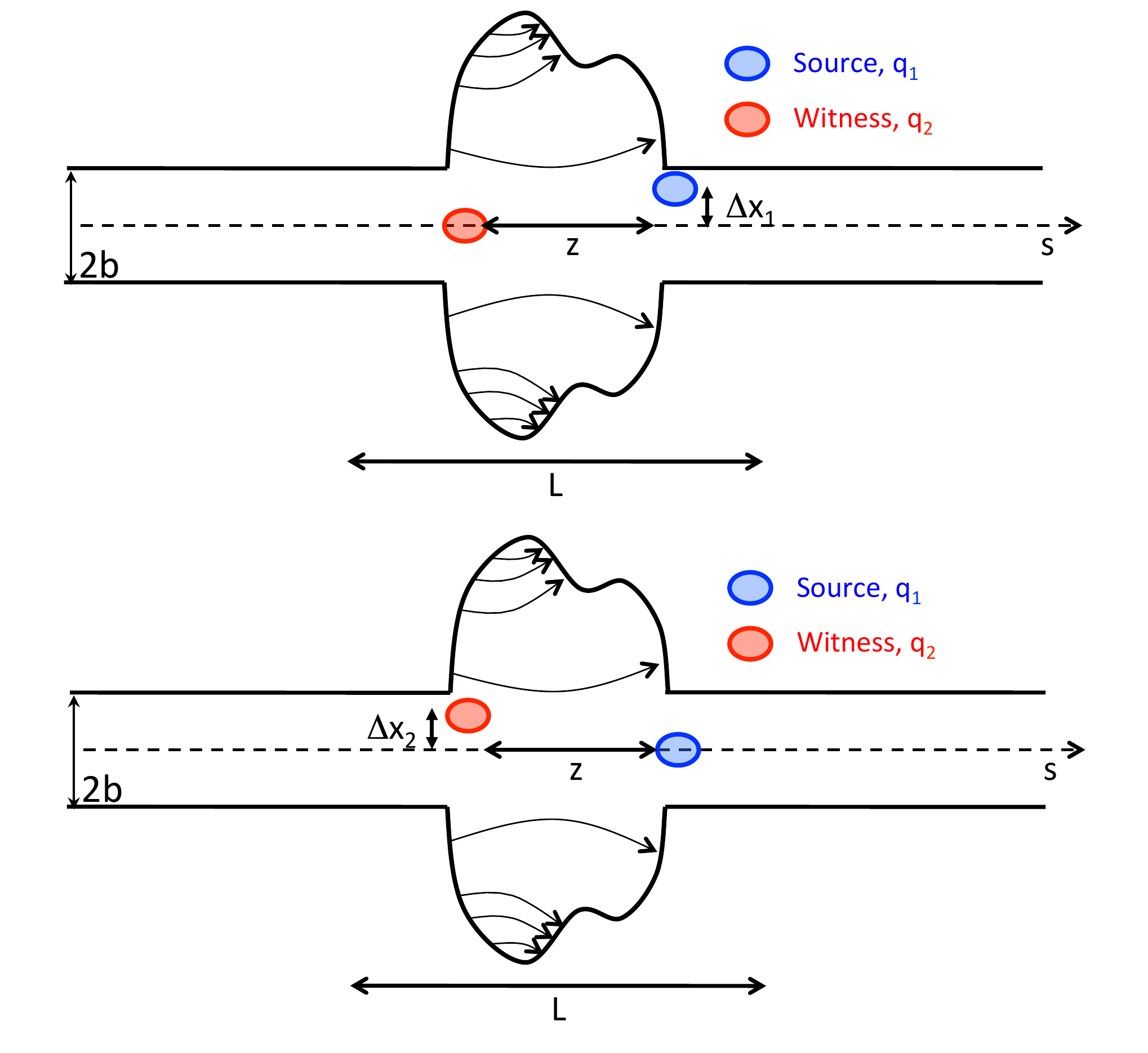}
\caption{Wake field from a source particle potentially affecting a witness travelling at distance $z$ behind the source: dipolar (top) and quadrupolar (bottom).}
\label{sketch-waketr}
\end{center}
\end{figure}

We can start from the same system we have used in the previous section to introduce the concept of longitudinal wake function. We consider two ultra-relativistic charged particles, $q_1$ and $q_2$, going through an accelerator structure, with the trailing (witness) particle at a distance $z$ from the leading (source) one. In an axisymmetric structure (or simply with a top--bottom and left--right symmetry) a source particle travelling on axis cannot induce net transverse forces on a witness particle also following on axis. A symmetry breaking has to be introduced to drive transverse effects, and at the first order there are two options, i.e. offset the source or the witness (see Fig.~\ref{sketch-waketr}). The {\em transverse (horizontal or vertical) dipolar wake function} associated with a certain accelerator object (able to create a wake field) is defined as the integrated transverse force from an offset source ($q_2\cdot [\vec{E}(s,z) + \vec{v}\times \vec{B}(s,z)]_{x,y}$) acting on the witness particle along the effective length of the object, normalized by the source and witness charges and by the offset of the source charge, $\Delta x_1$ or $\Delta y_1$ (see Fig.~\ref{sketch-waketr}, top):
\begin{equation}
\begin{array}{l}
\displaystyle W_{Dx}(z)=-\frac{\int_0^L \left[\vec{E}(s,z) + \vec{v}\times\vec{B}(s,z)\right]_x \, {\rm d}s}{q_1\Delta x_1} = -\left(\frac{E_0}{q_1q_2}\right)\frac{\Delta x'_2}{\Delta x_1},\\[5mm]
\displaystyle W_{Dy}(z)=-\frac{\int_0^L \left[\vec{E}(s,z) + \vec{v}\times\vec{B}(s,z)\right]_y \, {\rm d}s}{q_1\Delta y_1} = -\left(\frac{E_0}{q_1q_2}\right)\frac{\Delta y'_2}{\Delta y_1}.
\end{array}
\label{longwake2}
\end{equation}

The {\em transverse (horizontal or vertical) quadrupolar wake function} associated with a certain accelerator object (able to create a wake field) is defined as the integrated transverse force from an on-axis source ($q_2\cdot [\vec{E}(s,z) + \vec{v}\times \vec{B}(s,z)]_{x,y}$) acting on an offset witness particle along the effective length of the object, normalized by the source and witness charges and by the offset of the witness charge, $\Delta x_2$ or $\Delta y_2$ (see Fig.~\ref{sketch-waketr}, bottom):
\begin{equation}
\begin{array}{l}
\displaystyle W_{Qx}(z)=-\frac{\int_0^L \left[\vec{E}(s,z) + \vec{v}\times\vec{B}(s,z)\right]_x \, {\rm d}s}{q_1\Delta x_2} = -\left(\frac{E_0}{q_1q_2}\right)\frac{\Delta x'_2}{\Delta x_2},\\[5mm]
\displaystyle W_{Qy}(z)=-\frac{\int_0^L \left[\vec{E}(s,z) + \vec{v}\times\vec{B}(s,z)\right]_y \, {\rm d}s}{q_1\Delta y_2} = -\left(\frac{E_0}{q_1q_2}\right)\frac{\Delta y'_2}{\Delta y_2}.
\end{array}
\label{longwake3}
\end{equation}

For most objects of interest, it can be seen that the wake functions so defined do not depend on the source or witness offsets, provided the offsets are much smaller than the transverse size of the object. For larger offsets, coupling and/or higher order non-linear terms can become important and may need to be taken into account to describe correctly the particle dynamics. Both the dipolar and quadrupolar wake functions in $z=0$, $W_{Dx,Dy}(0)$ and $W_{Qx,Qy}(0)$, must vanish because for $z=0$ source and witness particles are travelling together and they can only mutually interact through space charge, which is not included in this framework. Besides, $W_{Dx,Dy}(0^-)$ is generally negative, because trailing particles tend to be deflected toward the source particle (offset and kick have the same sign). The sign of the quadrupolar wake functions in $0^-$ depends on the geometry and properties of the surrounding environment. As we also discussed for the longitudinal wake function, the condition that transverse wake functions must vanish for $z>0$ due to causality also holds.

The transverse wake function of an accelerator object is very useful for macroparticle models and simulations, because it relates source or witness perturbations to the associated kicks on trailing particles:
\begin{equation}
\left\{\begin{array}{l}
\displaystyle \Delta x'_2(z)=-\left(\frac{q_1q_2}{E_0}\right)\left[ W_{Dx}(z)\Delta x_1 + W_{Qx}(z)\Delta x_2\right], \\[5mm]
\displaystyle \Delta y'_2(z)=-\left(\frac{q_1q_2}{E_0}\right)\left[ W_{Dy}(z)\Delta y_1 + W_{Qy}(z)\Delta y_2\right].
\end{array}
\right.
\end{equation}

We can also describe the interaction as a transfer function in the frequency domain, which defines the {\em transverse beam coupling impedance (dipolar and quadrupolar)} of the object under study:
\begin{equation}
\left\{\begin{array}{l}
\displaystyle Z_{Dx,Dy}(\omega)={\rm i}\int_{-\infty}^{\infty}W_{Dx,Dy}(z)\exp\left(\frac{{\rm i}\omega z}{c}\right)\frac{{\rm d}z}{c},\\[5mm]
\displaystyle Z_{Qx,Qy}(\omega)={\rm i}\int_{-\infty}^{\infty}W_{Qx,Qy}(z)\exp\left(\frac{{\rm i}\omega z}{c}\right)\frac{{\rm d}z}{c}.
\end{array}
\right.
\label{transv-imp}
\end{equation}

\begin{figure}[htb]
\begin{center}
\includegraphics[trim = 10mm 10mm 10mm 10mm, clip, width=0.65\textwidth]{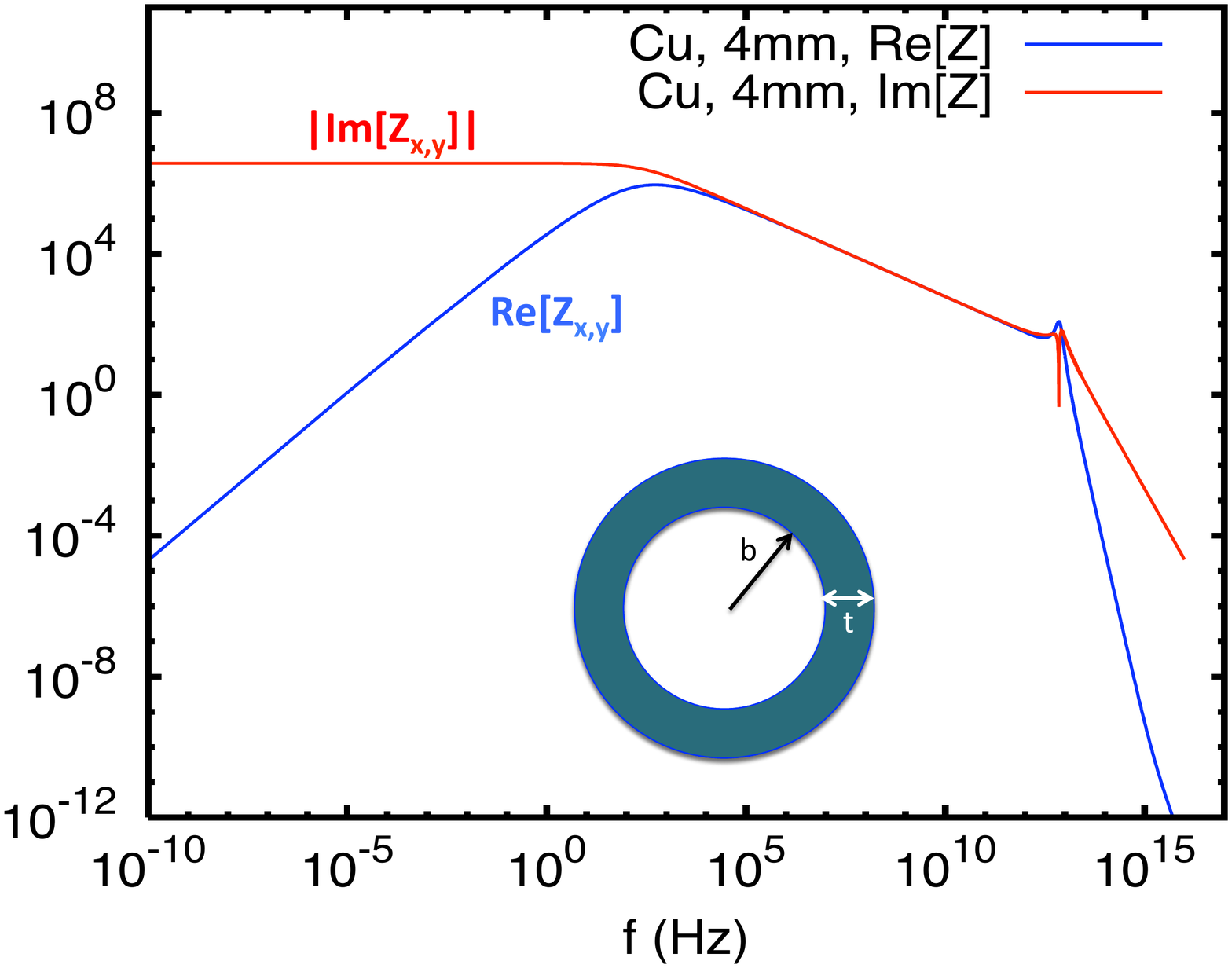}
\caption{Wall dipolar impedance for a cylindrical pipe of radius $b$ and thickness $t$, as illustrated. Courtesy of N.~Mounet.}
\label{reswall-imp}
\end{center}
\end{figure}

Similarly to the longitudinal plane, in the transverse plane typical wake/impedance pairs are also represented by resonators, which have a peaked structure in the frequency domain and are damped oscillations in the time domain. Another important example of transverse impedance is the wall impedance. Figure~\ref{reswall-imp} depicts the dipolar impedance spectrum for a simple cylindrical chamber with a wall of finite thickness $t$, finite conductivity $\sigma$ and radius $b$. This impedance extends over a very wide range of frequencies and exhibits different behaviours that can be intuitively understood as described in the following. At low frequencies, such that the penetration depth of the electromagnetic fields into the chamber is much larger than the wall thickness, i.e.~$\delta(\omega)=\sqrt{2/(\mu_0\sigma\omega)}\gg t$, the beam can only see the induced charges on the inner surface of the chamber, associated with a constant imaginary part of the impedance (betatron tune shift by extra defocusing) and basically zero real part (no losses). At intermediate frequencies, the electromagnetic interaction between the beam and the conducting pipe happens through a decreasing $\delta(\omega)$ and the impedance becomes about proportional to the penetration depth and decays like $\sqrt{\omega}$. At high frequency, a point is reached at which the electromagnetic fields can become trapped within the penetration depth, generating a resonant peak.

\begin{figure}[htb]
\begin{center}
\includegraphics[trim = 40mm 90mm 30mm 50mm, clip, width=1.0\textwidth]{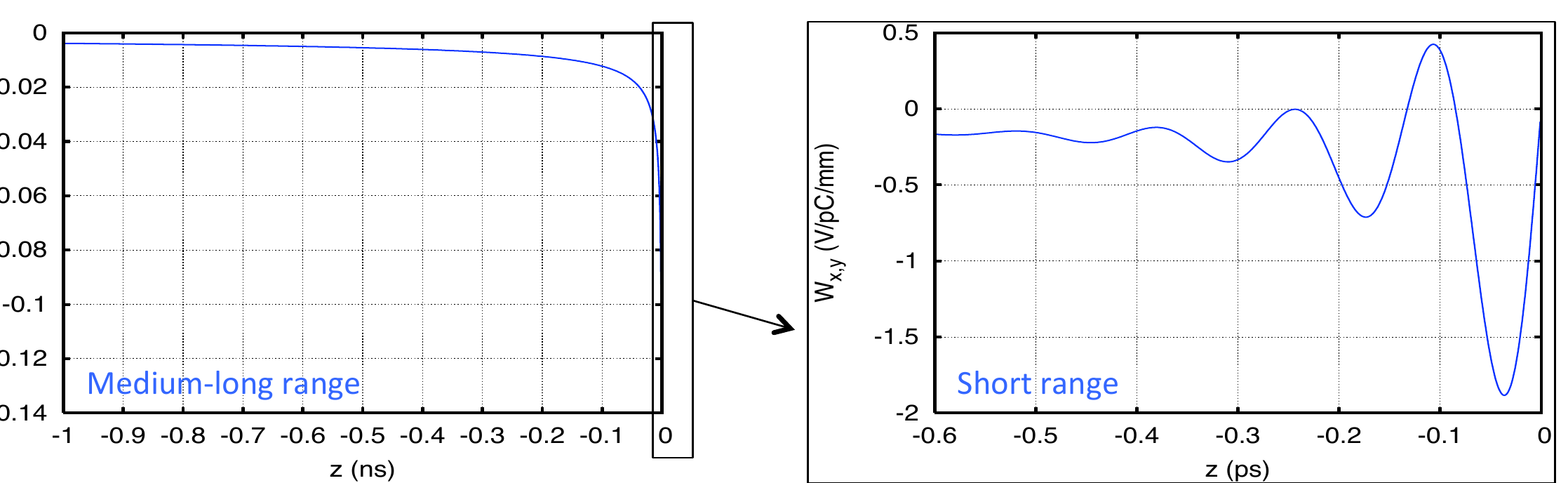}
\caption{Wall dipolar wake function for a cylindrical pipe of radius $b$ and thickness $t$. The short range has been zoomed in to highlight the oscillatory behaviour of the function at short distances from the source. Courtesy of N.~Mounet.}
\label{reswall-wake}
\end{center}
\end{figure}

Corresponding to the different frequency ranges, the wake function also exhibits different behaviours on different distance ranges from the source charge. In the medium--long range (usually coupled bunch and/or multiturn), the wake function is characterized by a monotonic decay slowly converging to zero, while in the short range (typically responsible for single-bunch effects) the high-frequency resonance peak dominates and a damped oscillatory behaviour is found. The wall wake function is displayed in Fig.~\ref{reswall-wake}, in which the short-range part has also been zoomed in order to highlight the different behaviour. The switch between the two regimes obviously depends on the frequency at which the peak is actually located.

It can be demonstrated that the transverse impedance of a resistive wall is inversely proportional to $b^3$, while its longitudinal counterpart is inversely proportional to $b$. That is why the transverse effects due to a resistive wall are in general more severe than the corresponding longitudinal ones. In particular, the transverse resistive wall impedance is responsible for coupled bunch instabilities and determines which damping time a feedback system needs in order to be able to efficiently counteract this effect. Furthermore, resistive wall effects become especially important in machines with low-emittance beams, which have chambers with very small radii and might require coatings with low-conductivity materials to avoid other effects.

\subsection{The rigid bunch instability}
Repeating the same procedure used for the longitudinal plane, to study instabilities in the transverse plane, the effect of wake fields must be formally introduced in the equations of transverse motion of the beam particles. Using the concepts introduced at the beginning of this section, we can write the equations of the transverse motion of any single particle in the witness slice $\lambda(z) \, {\rm d}z$ under the effect of the focusing force from the external magnets and that associated with a distributed wake, which can extend to several turns:
\begin{equation}
\left\{\begin{array}{l}
\displaystyle \frac{{\rm d}^2x}{{\rm d}s^2} + K_x(s)x = -\frac{e^2}{m_0c^2\gamma C}\cdot\\[3mm]
\displaystyle\;\;\;\;\;\;\cdot\sum_{k=0}^{\infty}\int_{z}^{\infty}\lambda(z'+kC,s)\left[\langle x \rangle(z'+kC,s)W_{Dx}(z-z'-kC) + x W_{Qx}(z-z'-kC) \right] \, {\rm d}z',\\[8mm]
\displaystyle\frac{{\rm d}^2y}{{\rm d}s^2} + K_y(s)y = -\frac{e^2}{m_0c^2\gamma C}\cdot\\[3mm]
\displaystyle\;\;\;\;\;\;\cdot\sum_{k=0}^{\infty}\int_{z}^{\infty}\lambda(z'+kC,s)\left[\langle y \rangle(z'+kC,s)W_{Dy}(z-z'-kC) + y W_{Qy}(z-z'-kC) \right] \, {\rm d}z'.
\end{array}
\right.
\label{transv-motion}
\end{equation}
\vspace*{0cm}

Here $K_{x,y}(s)$ are the external focusing coefficients as from Hill's equation and $\langle x, y\rangle(z,s)$ represent the average $x$ and $y$ across the ${\rm d}N_b(z)=\lambda(z,s) \, {\rm d}z$ particles contained in the thin slice ${\rm d}z$. Equation (\ref{transv-motion}) is very general and can be used in macroparticle tracking programs, which solve it for each macroparticle, determining self consistently the evolution of $\lambda(z,s)$ as well as of $\langle x, y\rangle(z,s)$. As in the longitudinal case, all integrals and summations in the above equations can be formally extended from $-\infty$, as the wake functions vanish for positive values of $z$ due to causality.

In the following, to illustrate the most basic mechanism of transverse instability, i.e. the {\em rigid bunch instability}, we will make use of a set of simplifying assumptions:
\begin{itemize}
\item The bunch is point-like, carries a charge $N_be$ and feels an external linear force (i.e. it would execute linear betatron oscillations in absence of the wake forces).
\item Longitudinal motion is neglected.
\item Smooth approximation (constant focusing).
\item Distributed wake, only dipolar and only in the $y$ direction.
\end{itemize}
The equation of motion of this one-particle beam can then be written
\begin{equation}
\displaystyle\frac{{\rm d}^2y}{{\rm d}s^2} + \left(\frac{\omega_{\beta}}{c}\right)^2y=-\left(\frac{N_be^2}{m_0c^2\gamma C}\right)\sum_{k=0}^{\infty}y(s+kC)W_{Dy}(-kC).
\end{equation}

Transforming into the frequency domain and applying the Poisson sum formula, we obtain
\[
\displaystyle\omega^2 - \omega_{\beta}^2 = \frac{N_be^2}{m_0\gamma C}\sum_{k=-\infty}^{\infty}\exp(ik\omega T_0)W_{Dy}(kC)=
\]
\begin{equation}
\displaystyle\;\;\;=-\frac{iN_be^2}{m_0\gamma CT_0}\sum_{p=-\infty}^{\infty}Z_y(p\omega_0+\omega) \;\; .
\end{equation}

Assuming that the effect of the wake results in a small deviation from the betatron tune, we can derive from the above equation a simple estimate of the real frequency shift (tune shift) and the imaginary frequency shift (growth/damping rate):
\begin{equation}
\begin{array}{l}
\displaystyle \frac{\mathrm{Re}(\omega-\omega_{\beta})}{\omega_0}=\Delta Q_y \approx \frac{N_be^2\langle\beta_y\rangle}{4\pi m_0\gamma c C}\sum_{p=-\infty}^{\infty}\mathrm{Im}\left[Z_{Dy}(p\omega_0+\omega_{\beta})\right], \\[6mm]
\displaystyle \mathrm{Im}(\omega-\omega_{\beta})=\tau^{-1}\approx -\frac{N_be^2\langle\beta_y\rangle}{2m_0\gamma C^2}\sum_{p=-\infty}^{\infty}\mathrm{Re}\left[Z_{Dy}(p\omega_0+\omega_{\beta})\right].
\end{array}
\label{rigid-eqs}
\end{equation}

With the given definition of the transverse impedance (i.e. including the imaginary unit in Eqs.~(\ref{transv-imp})), the tune shift is found to depend only on the imaginary part of the impedance, while the growth/damping rate depends only on its real part. It is interesting to note that the tune shift can also be expressed in the following well-known compact form:
\begin{equation}
\Delta Q_y = \frac{1}{4\pi}\left[\langle\beta_y\rangle \frac{eI_b\mathrm{Im}(Z_{Dy}^{\mathrm{eff}})}{E} \right]\rightarrow\frac{1}{4\pi}\oint\beta_y(s)\Delta k_y(s) \, {\rm d}s,
\end{equation}

\noindent where $\Delta k_y(s)$ is the distributed quadrupolar error due to the wake, which can be written as the relative energy kick from the wake $\Delta E_y/E$ divided by the circumference $C$. Furthermore, besides the tune shift, the presence of the wake introduces an imaginary part of the betatron frequency shift, which, if positive, can result in a beam instability. In particular, the summation in the second of the equations~(\ref{rigid-eqs}) can be positive or negative, because $\mathrm{Re}[Z_{Dy}(\omega)]$ is an odd function. Unlike the case of the Robinson instability in the longitudinal plane, here the sign of the imaginary frequency shift is solely determined by the sign of this summation. In a first noteworthy case, similarly to what we discussed in the longitudinal case, we can assume the transverse impedance to be peaked at a frequency $\omega_r$ close to $h\omega_0$ (e.g. RF cavity fundamental mode or a HOM). If we define the tune $Q_y=n_y + \Delta_{\beta y}$ with $-0.5<\Delta_{\beta y}<0.5$ and we make use of the property of the real part of the impedance to be an odd function of $\omega$, we can easily reduce the summation at the right-hand side of the second of the equations~(\ref{rigid-eqs}) to the sum of its two leading terms alone ($p$ such that $p+n_y=h$):
\begin{equation}
\tau^{-1}\approx -\frac{N_be^2\langle\beta_y\rangle}{2m_0\gamma C^2}\left(\mathrm{Re}[Z_{Dy}(h\omega_0+\Delta_{\beta y}\omega_0)] - \mathrm{Re}[Z_{Dy}(h\omega_0-\Delta_{\beta y}\omega_0)] \right).
\label{stab-res}
\end{equation}

\begin{figure}[htb]
\begin{center}
\includegraphics[trim =20mm 50mm 20mm 40mm, clip, width=0.9\textwidth]{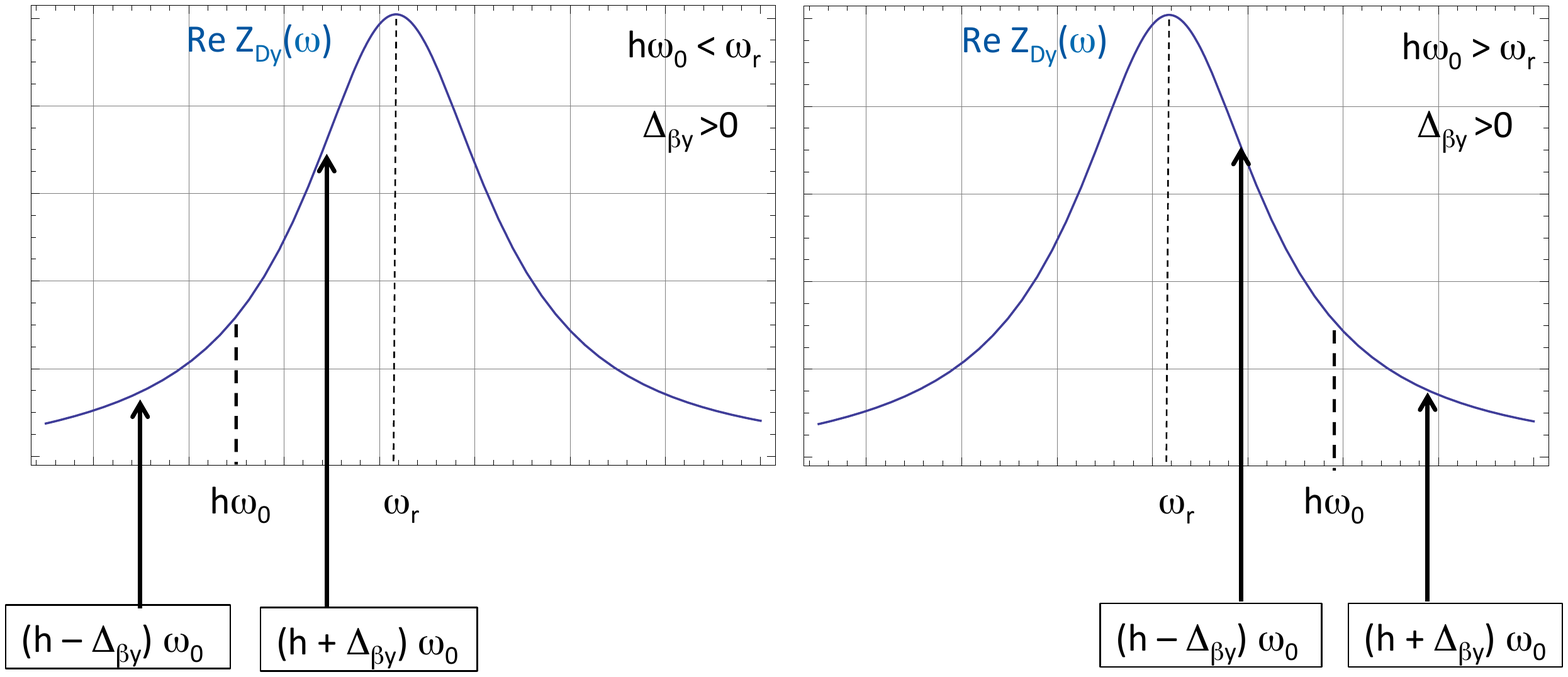}\\
{\small (a) \hspace{14pc} (b)}
\caption{Sketch of the two possible situations for the rigid bunch instability with resonator impedance}
\label{rigid-sketch}
\end{center}
\end{figure}

Figure \ref{rigid-sketch} illustrates the two possible situations that can occur, corresponding to the case of positive $\Delta_{\beta y}$, i.e. tune below the half integer. If $\omega_r>h\omega_0$ (Fig. \ref{rigid-sketch}(a)), the right-hand side of Eq.~(\ref{stab-res}) is negative and the bunch will be stable. Conversely, if $\omega_r<h\omega_0$
(Fig. \ref{rigid-sketch}(b)), the right-hand side of Eq.~(\ref{stab-res}) is positive, entailing an instability.
Obviously, the situation is reversed when the tune is above the half integer ($\Delta_{\beta y}<0$).

\begin{figure}[htb]
\begin{center}
\includegraphics[trim =60mm 0mm 0mm 0mm, clip, width=0.9\textwidth]{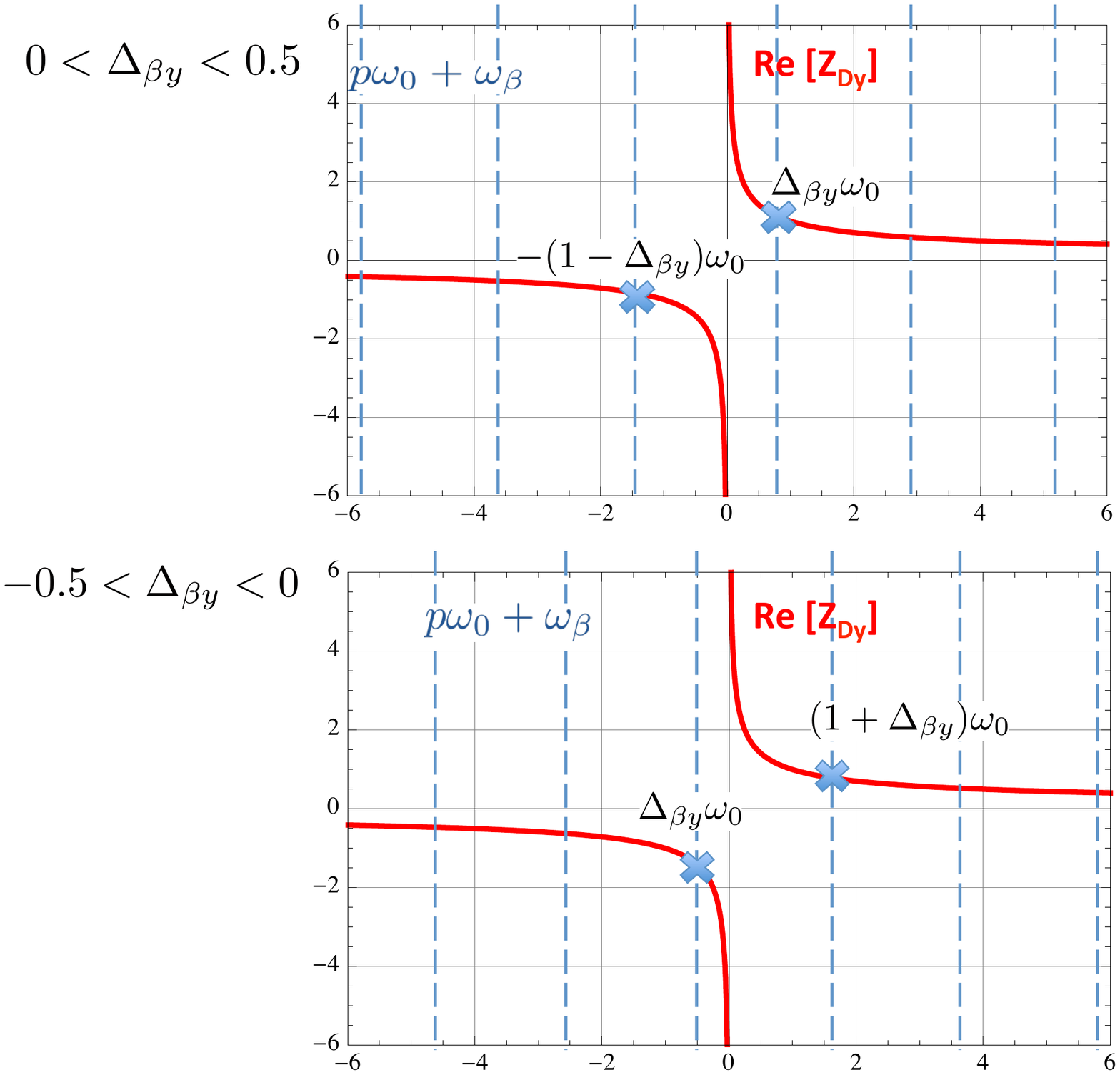}
\caption{Sketch of the two possible situations for the rigid bunch instability with resistive wall impedance}
\label{rigid-sketch-res}
\end{center}
\end{figure}

Another interesting case is when the impedance is of resistive wall type, i.e. strongly peaked in the very low frequency range (diverging for $\omega\rightarrow 0$ in the approximation of thick wall). Then we can distinguish the two situations depicted in Fig.~\ref{rigid-sketch-res}, i.e.~fractional tune below or above the half integer. In the former case, the two leading terms of the summation can be expressed as
\begin{equation}
\tau^{-1}\approx -\frac{N_be^2\langle\beta_y\rangle}{2m_0\gamma C^2}\left(\mathrm{Re}[Z_{Dy}(\Delta_{\beta y}\omega_0)] - \mathrm{Re}[Z_{Dy}((1-\Delta_{\beta y})\omega_0)] \right),
\label{stab-resb}
\end{equation}
\noindent which is negative (see Fig.~\ref{rigid-sketch-res}, top plot), ensuring beam stability. In the latter case, i.e. for fractional tune above the half integer, we obtain
\begin{equation}
\tau^{-1}\approx -\frac{N_be^2\langle\beta_y\rangle}{2m_0\gamma C^2}\left(\mathrm{Re}[Z_{Dy}((1+\Delta_{\beta y})\omega_0)] -\mathrm{Re}[Z_{Dy}(-\Delta_{\beta y}\omega_0)] \right).
\label{stab-res2}
\end{equation}

As is visible from Fig.~\ref{rigid-sketch-res}, bottom plot, this is positive, leading in any case to an instability. This is the reason why most of the running machines are usually operated with a fractional part of the tunes below 0.5, although, in practice, tunes above the half integer can be used, if the resistive wall instability is Landau damped or efficiently suppressed with a feedback system.

\subsection{Strong head--tail instability and transverse-mode coupling}
Making now one step further in the description of the mechanisms of transverse instability, the case of the strong head--tail instability, also called transverse mode coupling instability (TMCI), can be illustrated by means of a simple two-particle model. It is assumed that the beam consists of two macroparticles, each having a charge of $\Np_be/2$. They perform synchrotron oscillations of the same frequency and amplitude, but with opposite phase. During half a synchrotron period $T_s=2\pi/\omega_s$, particle with index 1 is leading and thus performing free betatron oscillations, while the trailing particle indexed 2 feels the wake field generated by particle 1. Thus, for $0 < s < \pi c/\omega_s$, assuming zero chromaticity, and therefore no dependence of the frequency of the transverse oscillation on the longitudinal parameters, the equations of motion for the two macroparticles are simply written as
\begin{subequations}
	\begin{align}
	y_1''+\left(\frac{\omega_\beta}{c}\right)^2 y_1 &= 0 	\label{EQ:2ParticleEquationOfMotionLeading}, \\
	y_2''+\left(\frac{\omega_\beta}{c}\right)^2 y_2 &= \left(\frac{e^2}{m_0c^2}\right)\frac{\Np_b \wake_0}{2 \gammarel C}\,y_1(s),
	\label{EQ:2ParticleEquationOfMotionTrailing}
	\end{align}
\label{EQ:2ParticleModelEquationsOfMotion}	
\end{subequations}
where $y_1$ denotes the vertical position of particle 1 and $y_2$ the position of particle 2, and the focusing term in Hill's equation has been written as $K_y=\left({\omega_\beta}/{c}\right)^2$ with $\omega_\beta$ denoting the (vertical) betatron frequency. Note that it is assumed here that the wake field $\wake_0$ (integrated over the machine circumference $C$) is constant but vanishes before the beam passage in the consecutive turn. This corresponds practically to the case of a broadband impedance.

The stability of the two-particle system is analysed in the following. The solution for the free betatron oscillation of Eq.~\eqref{EQ:2ParticleEquationOfMotionLeading} can be written as
\begin{equation}
\tilde y_1(s)=\tilde y_1 (0) \exp{\left(\frac{-{\rm i} \omega_\beta s}{c}\right)},
\end{equation}
where the complex phasor $\tilde y_{1}(s)$
\begin{equation}
\tilde y_{1}(s)=y_{1}(s)+{\rm i}\frac{c}{\omega_\beta}y'_{1}(s)
\end{equation}
has been introduced. Inserting the solution for $\tilde y_1(s)$ into Eq.~\eqref{EQ:2ParticleEquationOfMotionTrailing} leads to the solution for $\tilde y_2(s)$
\begin{equation}
\tilde y_2(s)={\tilde y_2 (0) \exp\!{\left(\!-\frac{{\rm i} \omega_\beta s}{c}\right)}} +{{\rm i} \frac{\Np_b e^2 \wake_0}{4 m_0 \gammarel c C \omega_\beta} \left[\frac{c}{\omega_\beta}\tilde y_1^*(0) \sin{\left( \frac{\omega_\beta s}{c}\right)\!+ \tilde y_1 (0)\,s\exp\!{\left(\!-\frac{{\rm i} \omega_\beta s}{c}\right)}} \right]},
\label{EQ:PositionTrailingParticle}
\end{equation}
which consists of the free betatron oscillation term and a driven oscillation term. For the further analysis, the position of the two particles is evaluated at $s=\pi c/\omega_s$, i.e.~after half the synchrotron period. Since the betatron frequency is typically much larger than the synchrotron frequency, i.e.~$\omega_\beta\gg\omega_s$, the second term on the right-hand side of Eq.~\eqref{EQ:PositionTrailingParticle} is small compared to the last term. Thus, the solutions of the equations of motion can be written in matrix form:
\begin{equation}
\left(\begin{array}{c}\tilde y_1 \\\tilde y_2 \end{array}\right)_{s=\pi c/\omega_s}=\exp{\left(\!-\frac{{\rm i} \pi \omega_\beta}{\omega_s}\right)}\cdot\left(\begin{array}{cc}1 & 0 \\ {\rm i} \Upsilon & 1 \end{array}\right)\cdot\left(\begin{array}{c}\tilde y_1 \\\tilde y_2 \end{array}\right)_{s=0},
\end{equation}
where the positive dimensionless parameter $\Upsilon$ has been defined as
\begin{equation}
\Upsilon=\frac{\pi \Np_b e^2 \wake_0}{4 m_0 \gammarel C \omega_\beta \omega_s}.
\end{equation}

During the second half of the synchrotron period, i.e.~$\pi c/\omega_s < s < 2\pi c/\omega_s$, the two particles exchange their roles and now particle 2 is leading, while particle 1 is feeling the wake. Thus, the equations of motion have to be exchanged and, by combining the transformations over the two half synchrotron periods, the transformation matrix for the full synchrotron period can be finally obtained as
\begin{subequations}
\begin{align}
\left(\begin{array}{c}\tilde y_1 \\\tilde y_2 \end{array}\right)_{s=2\pi c/\omega_s}&=\exp{\left(\!-\frac{{\rm i} 2\pi \omega_\beta}{\omega_s}\right)}\cdot\left(\begin{array}{cc}1 &  {\rm i} \Upsilon \\ 0 & 1 \end{array}\right)\cdot\left(\begin{array}{cc}1 & 0 \\  {\rm i} \Upsilon  & 1 \end{array}\right)\cdot\left(\begin{array}{c}\tilde y_1 \\\tilde y_2 \end{array}\right)_{s=0}\\
&=\exp{\left(\!-\frac{{\rm i} 2\pi \omega_\beta}{\omega_s}\right)}\cdot\left(\begin{array}{cc}1 - \Upsilon^2 & {\rm i}\Upsilon \\ {\rm i}\Upsilon & 1 \end{array}\right)\cdot\left(\begin{array}{c}\tilde y_1 \\\tilde y_2 \end{array}\right)_{s=0}.
\end{align}
\end{subequations}
The stability of the system is determined by the eigenvalues of the transformation matrix. The characteristic equation for the two eigenvalues $\lambda_{\pm}$ yields
\begin{equation}
\lambda_{\pm}=\left(1-\Upsilon^2/2\right)\pm\sqrt{\left(1-\Upsilon^2/2\right)^2-1}.
\label{EQ:2ParticleModelEigenvalueEquation}
\end{equation}
Since the product of the two eigenvalues is equal to 1, the condition for stability requires that they should be purely imaginary exponentials, i.e.
\begin{equation}
\lambda_{+}\cdot\lambda_{-}=1 ~ \Rightarrow ~ \lambda_{1,2}=\exp{(\pm {\rm i}\upsilon)}.
\end{equation}

Inserting this back into Eq.~\eqref{EQ:2ParticleModelEigenvalueEquation} yields finally
\begin{equation}
\lambda_{+}+\lambda_{-}=2-\Upsilon^2 ~ \Rightarrow ~ \sin{\left(\frac{\upsilon}{2} \right)}=\frac{\Upsilon}{2}.
\end{equation}

Stability requires that $\upsilon$ should be real, which in turn is satisfied only if $\Upsilon\leq 2$. Therefore, the condition for stability written in terms of wake, machine and beam parameters reads
\begin{equation}
\Upsilon = \frac{\pi \Np_b e^2 \wake_0}{4m_0\gammarel C \omega_\beta\omega_s}\leq2.
\end{equation}

The threshold intensity for the onset of the strong head--tail instability in the two-particle model is thus obtained as
\begin{equation}
\Np_{b,\text{thr}}=\frac{8}{\pi e^2}\frac{p_0 \omega_s}{\beta_y}\left(\frac{C}{\wake_0}\right).
\label{EQ:TMCIthreshold2ParticleModel}
\end{equation}

From Eq.~(\ref{EQ:TMCIthreshold2ParticleModel}), we can deduce the following main features of this instability. The intensity threshold:
\begin{itemize}
\item is proportional to the momentum $p_o$, i.e.~bunches with higher energy are more stable;
\item scales proportionally with the synchrotron frequency $\omega_s$, i.e.~faster synchrotron motion helps increasing the stability range;
\item is inversely proportional to the beta function at the location of the impedance source, which is expected because the strength of a kick is always proportional to the beta function at the kick location;
\item is inversely proportional to the integrated wake field around the ring per unit length $\wake_0/C$, which means that a larger wake will decrease the intensity threshold.
\end{itemize}
The evolution of the centre of charge of the beam in the two-particle model is obtained by the sum of $\tilde y_1 + \tilde y_2$, which is found as
\begin{equation}
\begin{split}
\left(\tilde y_1 + \tilde y_2\right)(s) &= \exp\left[-{\rm i}\left(\omega_\beta \mp \frac{\upsilon \omega_s}{2 \pi} \right)\frac{s}{c}\right] \sum_{m=-\infty}^{\infty}C_m \exp{\left(-\frac{{\rm i} m \omega_s s}{c} \right)},\\
C_m &= 2 {\rm i} \Upsilon \frac{1\pm (-1)^m}{(2\pi m \mp \upsilon)^2}\left( 1\mp e^{\pm {\rm i} \upsilon /2}\right)
\end{split}
\end{equation}
with the amplitude coefficients $C_m$ for the oscillation modes with the mode number $m$. The oscillation frequencies of these
modes are given by
\begin{equation}
\begin{cases}
\Omega_+ = \omega_\beta+m\omega_s+\upsilon\omega_s/2\pi,\quad m~\text{even},\\
\Omega_- = \omega_\beta+m\omega_s-\upsilon\omega_s/2\pi,\quad m~\text{odd}.
\end{cases}
\end{equation}

\begin{figure}[t]
\begin{center}
\includegraphics[trim = 0mm 0mm 0mm 0mm, clip, width=0.56\textwidth]{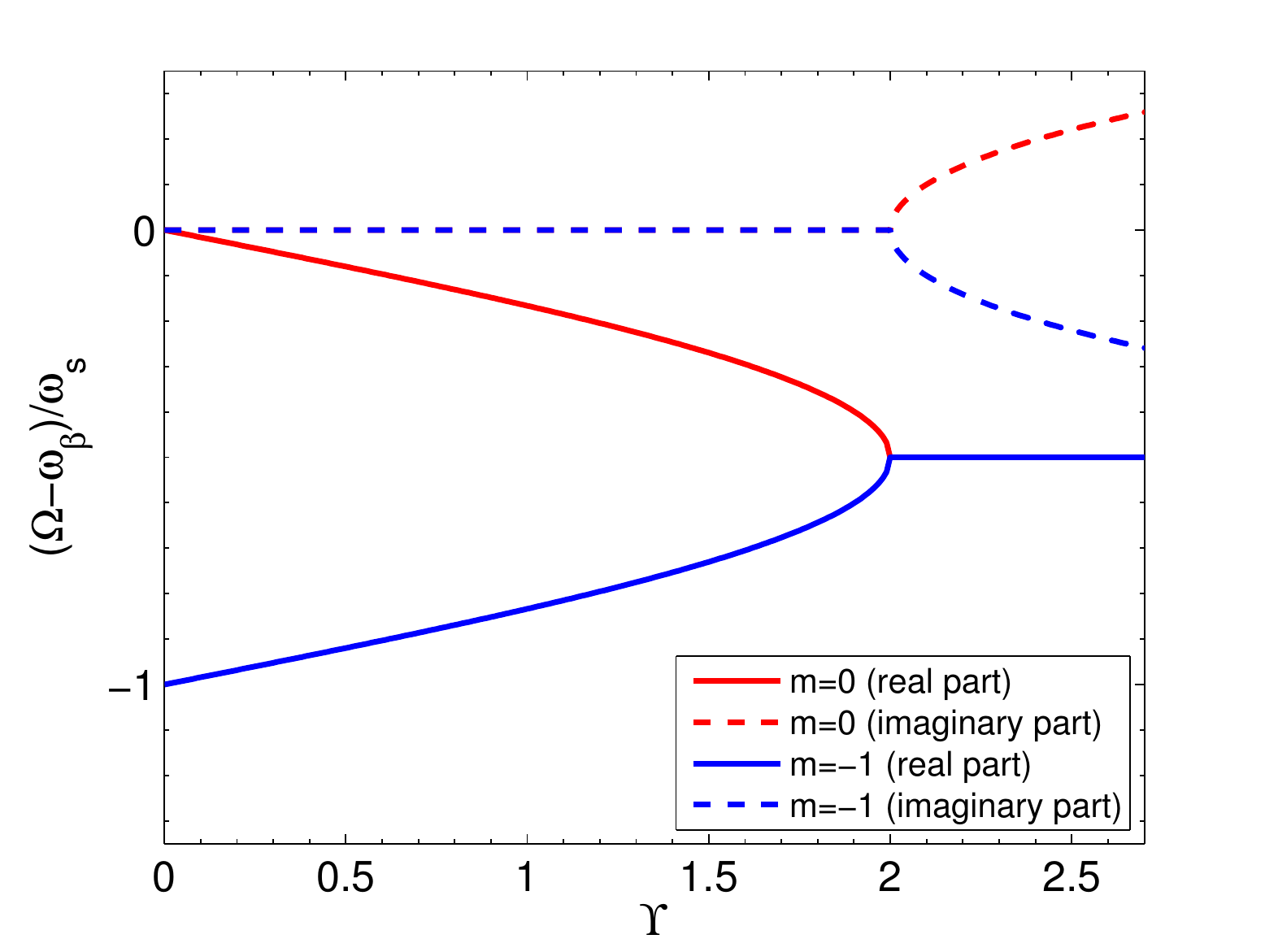}
\caption{Frequency spectrum of the centre of charge motion as a function of the parameter $\Upsilon$ as predicted by the two-particle model.}
\label{FIG:ModeCoupling2ParticleModel}
\end{center}
\end{figure}

Thus, as a function of the beam intensity the modes are shifting in frequency through the dependence on $\upsilon$. Figure~\ref{FIG:ModeCoupling2ParticleModel} shows the frequencies of these modes for $m=0$ and $m=-1$ as a function of $\Upsilon$. The two modes merge at $\Upsilon=2$ and the oscillation frequency becomes imaginary, i.e.~the beam becomes unstable and exhibits exponential growth. This is illustrated by plotting also the imaginary part of the oscillation frequencies. The strong head--tail instability is therefore also called TMCI.

\begin{figure}[htb]
\begin{center}
\includegraphics[trim =35mm 50mm 45mm 25mm, clip, width=0.85\textwidth]{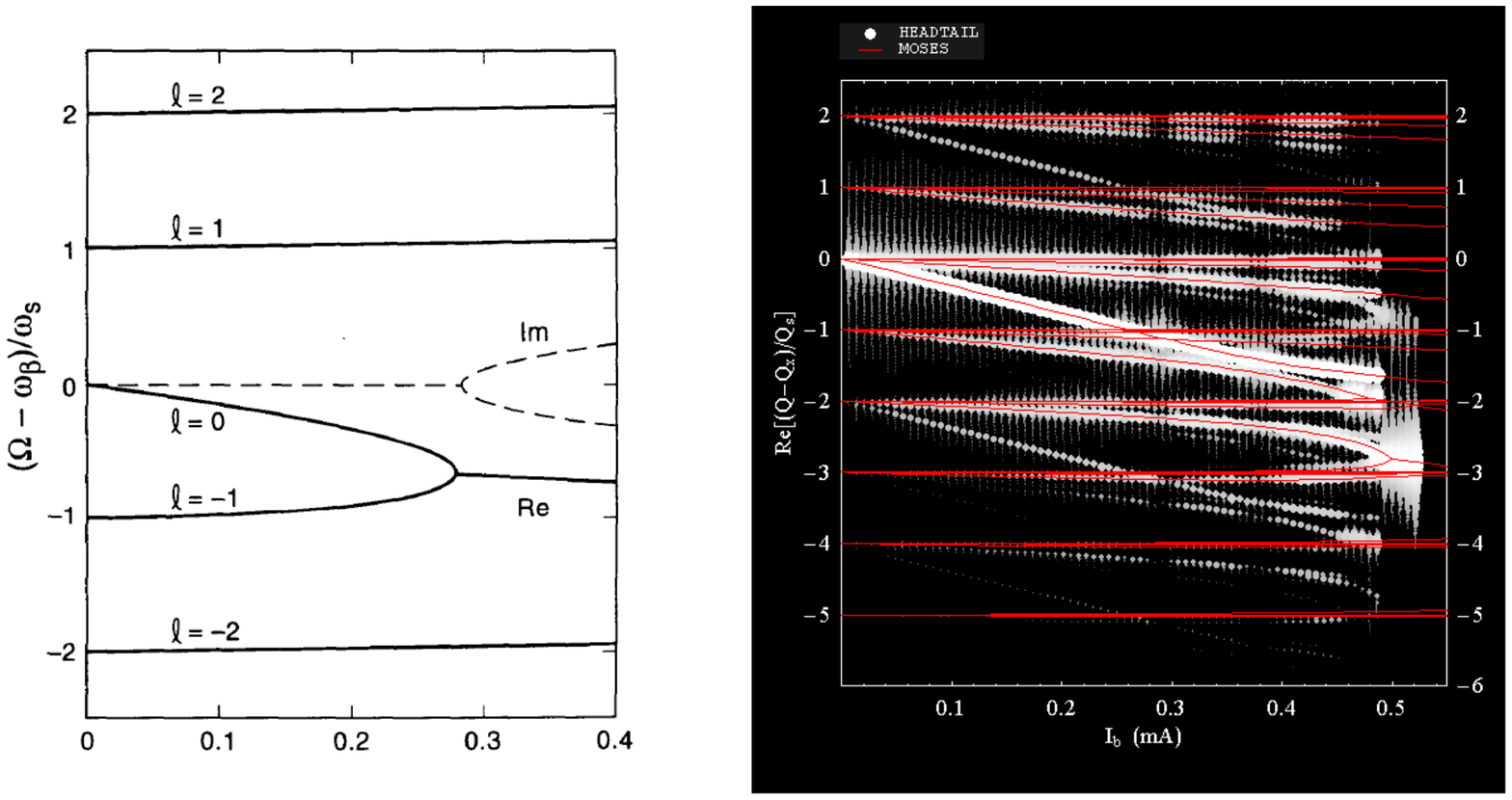}
\caption{Examples of mode shifts for a short bunch ((a), courtesy of A.~Chao) and for a long bunch ((b), courtesy of B.~Salvant). The left plot contains both the results of an analytical calculation (solid red lines) and those from macroparticle simulations (white lines).}
\label{modes}
\end{center}
\end{figure}

Beyond the two-particle model, several analytical formalisms have been developed for describing the TMCI. Good agreement between the different approaches is obtained when assuming a broadband resonator $\BBimpedanceVertical$ as driving impedance,
\begin{equation}
\BBimpedanceVertical (\omega)=\frac{\ResonatorFrequency}{\omega}\frac{\shuntImpedance}{\displaystyle 1+{\rm i}\QualityFactor \left(\frac{\omega}{\ResonatorFrequency} - \frac{\ResonatorFrequency}{\omega} \right)},
\end{equation}
where $\ResonatorFrequency$ is the resonance angular frequency,  $\QualityFactor$ is the resonator quality factor and $\shuntImpedance$ is the resonator shunt impedance (in $\Omega$/m). In the long bunch regime, i.e.~$\blength>\pi/\ResonatorFrequency$, the TMCI threshold can be obtained for example from the quasi-coasting beam approach using the peak values of bunch current and momentum spread, which yields
\begin{equation}
N_{b,\mathrm{thr}}^{\scriptscriptstyle \text{TMC}}=\frac{16\sqrt{2}}{3\pi} \frac{C |\eta| \emitlong}{ \langle\beta_y\rangle e c}\frac{\ResonatorFrequency}{|\BBimpedanceVertical|}\left(1+\frac{ Q'_y\,\omegarev }{\eta\,\ResonatorFrequency} \right), \label{EQ:TMCIthreshold}
\end{equation}
where $C$ is the machine circumference, $|\BBimpedanceVertical|$ is the peak value of the broadband resonator impedance and $\omegarev$ is the angular revolution frequency. Note that in comparison to the instability threshold obtained with the two-particle model in Eq.~\eqref{EQ:TMCIthreshold2ParticleModel}, the TMCI intensity threshold depends here in addition to the synchrotron tune (through the slip factor $\eta$) also on the longitudinal emittance $\emitlong$. Furthermore, the threshold can be raised by operating the machine with positive (negative) chromaticity above (below) transition. For a real bunch under the effect of a generic impedance, modes usually exhibit a more complicated shift pattern, which can be calculated via the Vlasov equation or can be found through macroparticle simulations. Examples of more complicated mode shift pictures for a short bunch (a) and a long Super Proton Synchrotron (SPS) bunch (b) under the effect of a broadband impedance are shown in Fig.~\ref{modes}.

Apart from the TMCI, it can be demonstrated that, including a non-zero value of chromaticity in the previous analysis, the individual coherent modes of a single bunch are intrinsically unstable (head--tail instabilities), even below the threshold for which they couple and give rise to TMCI. In particular, the main mode (mode $m=0$, corresponding to the betatron frequency) is naturally unstable below transition with positive chromaticity or above transition with negative chromaticity. Correspondingly, all higher order modes, $m\geq 1$, are naturally unstable below transition with negative chromaticity or above transition with positive chromaticity. Since, in practice, the most dangerous mode for a bunch is mode $m=0$, which is associated with the fastest rise time, accelerators are always operated with settings such as to keep this mode stable, while the other modes, slowly unstable, are damped through other mechanisms. That is the reason why low-energy accelerators operating below transition energy do not need chromaticity correction and can operate with their natural chromaticity (usually negative) without having problems of beam stability. On the other hand, high-energy accelerators operating above transition energy need sextupoles to correct chromaticity and stabilize the otherwise unstable mode $m=0$. Accelerators crossing transition need to make a chromaticity jump upon transition crossing, such as to ensure that the conditions for stabilizing mode $m=0$ are fulfilled at all times during the cycle.

\section{Final remarks}
Although all the mechanisms for beam instability reviewed in this article might seem to define sharp instability boundaries and thin parameter ranges for the operation of accelerators, in real life beam stability is eased by some other mechanisms not included in the simple models analysed in this article:
\begin{itemize}
\item Spreads of the beam characteristic frequencies and the possible associated non-linearities have a natural stabilizing action through Landau damping. Examples are momentum spread and synchrotron frequency spread in the longitudinal plane, or chromaticity and amplitude detuning in the transverse plane.
\item Active feedback systems are routinely employed to control/suppress instabilities. The principle is that the onset of a beam coherent motion is detected through a pick up, which sends a signal to a kicker that acts back on the beam to damp the motion before it can cause any degradation. Most of the running accelerators rely on this type of device, which is especially efficient against coupled-bunch instabilities. For single-bunch effects, especially in machines operating with short bunches, bandwidth and power requirements can be very stringent, potentially putting a technological limit to the feasibility of the system.
\end{itemize}
Furthermore, nowadays there is also a constant effort to identify, monitor and control impedance sources in present or future machines. In particular, impedance localization and reduction techniques are applied to running accelerators as well as for the design of new accelerators to extend their performance reach or ensure a smooth operation with the desired (target) parameter sets.

\section*{Acknowledgements}
The author would like to thank H. Bartosik, G. Iadarola, K. Li, E. M\'{e}tral, N.~Mounet, B.~Salvant, R.~Tom\'{a}s and C. Zannini for their invaluable help and the input/material kindly provided for the preparation of this article.

\end{document}